\documentclass[fleqn,usenatbib]{mnras}

\usepackage{newtxtext,newtxmath}

\usepackage[T1]{fontenc}
\usepackage{ae,aecompl}

\newcommand{\ccf}{\mathbf{C}(v_i,t_j)}
\newcommand{\meanccf}{\left<\mathbf{C}(v_i)\right> }
\newcommand{\meanacf}{\left<\mathbf{A}(\delta v_i)\right>}

\usepackage{graphicx}	
\usepackage{amsmath}	

\title[Planet-star signal separation]{Separating planetary reflex Doppler shifts from stellar variability in the  wavelength domain}

\author[A Collier Cameron et al.]{
A. Collier Cameron,$^{1,20}$\thanks{Contact e-mail: \href{mailto:acc4@st-andrews.ac.uk}{acc4@st-andrews.ac.uk}}
E. B. Ford,$^{2,3,4,20}$
S. Shahaf,$^{5}$
S. Aigrain,$^{6}$
X. Dumusque,$^{7}$
\newauthor
R. D. Haywood,$^{8,18 \dagger}$
A. Mortier,$^{9,17}$
D. F. Phillips,$^{8}$
L. Buchhave,$^{10}$
M. Cecconi,$^{11}$
\newauthor
H. Cegla,$^{7, 19}$
R. Cosentino,$^{11}$
M. Cr{\'e}tignier,$^{7}$
A. Ghedina,$^{11}$
M. Gonz\'alez,$^{11}$
\newauthor
D. W. Latham,$^{8}$
M. Lodi,$^{11}$
M. L{\'o}pez-Morales,$^{8}$
G. Micela,$^{12}$
E. Molinari,$^{13}$
F. Pepe,$^{7}$
\newauthor
G. Piotto,$^{14}$
E. Poretti,$^{11}$
D. Queloz,$^{9}$
J. San Juan,$^{11}$
D. S{\'e}gransan,$^{7}$
A. Sozzetti,$^{15}$
\newauthor
A. Szentgyorgyi,$^{8}$
S. Thompson,$^{9}$
S. Udry,$^{7}$
C. Watson$^{16}$
\\
\\
(Affiliations are listed at the end of the paper)}

\date{Accepted 2021 May 2. Received 2021 April 28; in original form 2020 October 13}

\pubyear{2019}

\begin{document}
\label{firstpage}
\pagerange{\pageref{firstpage}--\pageref{lastpage}}
\maketitle

\begin{abstract}
Stellar magnetic activity produces time-varying distortions in the photospheric line profiles of solar-type stars. 
\textcolor{black}{These lead to}
systematic errors in high-precision radial-velocity measurements, which limit efforts to discover and measure the masses of low-mass exoplanets with orbital periods of more than a few tens of days.
We present a new data-driven method for separating Doppler shifts of dynamical origin from apparent velocity variations arising from variability-induced changes in the stellar spectrum. 
\textcolor{black}{We show that the autocorrelation function (ACF) of the cross-correlation function used to measure radial velocities is effectively invariant to translation.}
\textcolor{black}{By projecting the radial velocities on to 
a subspace labelled by the observation identifiers and
spanned by the amplitude coefficients of the ACF's principal components, we can isolate and subtract velocity perturbations caused by stellar magnetic activity.}
\textcolor{black}{
We test the method on a 5-year time sequence of 853 daily 15-minute observations} of the solar spectrum from the HARPS-N instrument and solar-telescope feed on the 3.58-m Telescopio Nazionale Galileo. 
\textcolor{black}{After removal of the activity signals, the heliocentric solar velocity residuals are found to be Gaussian and nearly uncorrelated.}
We inject synthetic low-mass planet signals with amplitude $K=40$~cm~s$^{-1}$ into the solar observations  at a wide range of orbital periods. 
\textcolor{black}{Projection into the orthogonal complement of the ACF subspace} isolates these signals effectively from solar activity signals. Their semi-amplitudes are recovered with a precision of \textcolor{black}{$\sim~6.6$~cm~s$^{-1}$}, opening the door to Doppler detection and characterization of terrestrial-mass planets around well-observed, bright main-sequence stars across a wide range of orbital periods. 
\end{abstract}

\begin{keywords}
methods: statistical -- planets and satellites: general -- 
Sun: photosphere -- techniques: radial velocities -- techniques: spectroscopic 
\end{keywords}



\section{Introduction}

For decades, Doppler spectroscopy has been one of the most productive methods to discover and characterize exoplanets.
Improvements in the precision, wavelength calibration and stability of high-resolution {\'e}chelle spectrographs has allowed exoplanet surveys to probe planets with radial velocity (RV) semi-amplitudes of just $\sim 1$ m~s$^{-1}$.  
New generations of spectrographs such as 
CARMENES \citet{2014SPIE.9147E..1FQ}, 
ESPRESSO \citep{Megevand2014}, 
EXPRES \citep{Jurgenson2016}, 
\textcolor{black}{HARPS-3 \citep{2016SPIE.9908E..6FT},}
HPF \citep{Ninan2018} and 
NEID \citep{Schwab2016} 
are being designed and commissioned with improved resolution, spectral coverage, wavelength calibration and stabilization systems \citep{,2017RNAAS...1...51W}. Recently, ESPRESSO has achieved 30 cm s$^{-1}$ precision per RV observation on Proxima Cen \citep{2020A&A...639A..77S}, and EXPRESS 58 cm s$^{-1}$ on HD 3651 \citep{2020AJ....160...67B}.

Even with present instruments, the ability of spectroscopic surveys to detect and characterize low-mass planets is often limited by stellar variability and the stability of the wavelength calibration, rather than photon noise or instrumental errors \citep[e.g.,][]{1997ApJ...485..319S,2001A&A...379..279Q,2014MNRAS.443.2517H}.  
The purpose of the present study is to devise a practical new approach to measuring stellar radial velocities in a way that mitigates the errors due to line-shape changes caused by stellar variability. To achieve this, we make use of the fact that changes in the shape of spectral lines may influence the apparent radial velocity, but changes in the range rate (the first derivative with respect to time of the distance from the star's centre to the solar-system barycentre) induce only a shift and do not affect the line shape or depth. Related approaches exploiting profile-shape changes of even and odd character to disentangle shifts from activity have been published recently by \citet{2020MNRAS.491.4131Z} and \citet{2020arXiv200514083H}\textcolor{black}{, while \citet{2020arXiv201100003D} have employed a neural-network machine-learning to relating activity-related radial-velocity shifts to CCF profile-shape changes in the same solar dataset examined here.}

One feature of the method we present is that it makes use of existing data products, i.e. the cross-correlation functions (CCFs) between the observed spectra and a digital mask (see Section~\ref{sec:CCF}), and the radial velocities derived from them. 
The cross-correlation function $C(v,t)$ is a function of both barycentric velocity $v$ and time $t$. The temporal variability of the CCF includes both Doppler shifts and line shape changes.
The derivatives of the CCF with respect to velocity (e.g., $C'(v,t)$, $C''(v,t)$) are also functions of velocity and time.

Our long-term goal is to improve the detection sensitivity and robustness of spectroscopic planet searches.  
As a specific objective towards that goal, we aim to devise efficient, shift-invariant metrics that can contribute to characterizing the detailed line-profile shape at each epoch. 
We describe the building blocks for our new method in Section~\ref{sec:CCF}.  
Then, in Section~\ref{sec:scalpels}, we propose a new algorithm to compute such metrics, based on a novel combination of existing data products and techniques.  
We apply these metrics as predictors for the contribution of stellar variability to the apparent Doppler shift and infer a cleaned velocity time series.  
In Section~\ref{sec:injtests}, we verify and validate the method based on injection and recovery tests using solar observations.   
While our data-driven method makes no assumptions about the physical origin of the stellar variability, the tests in this paper focus on magnetic activity of the Sun.  
Finally, we discuss the implications of our work for future spectroscopic planet surveys in Section~\ref{sec:discuss}.

\begin{figure*}
    \includegraphics[width=2\columnwidth]{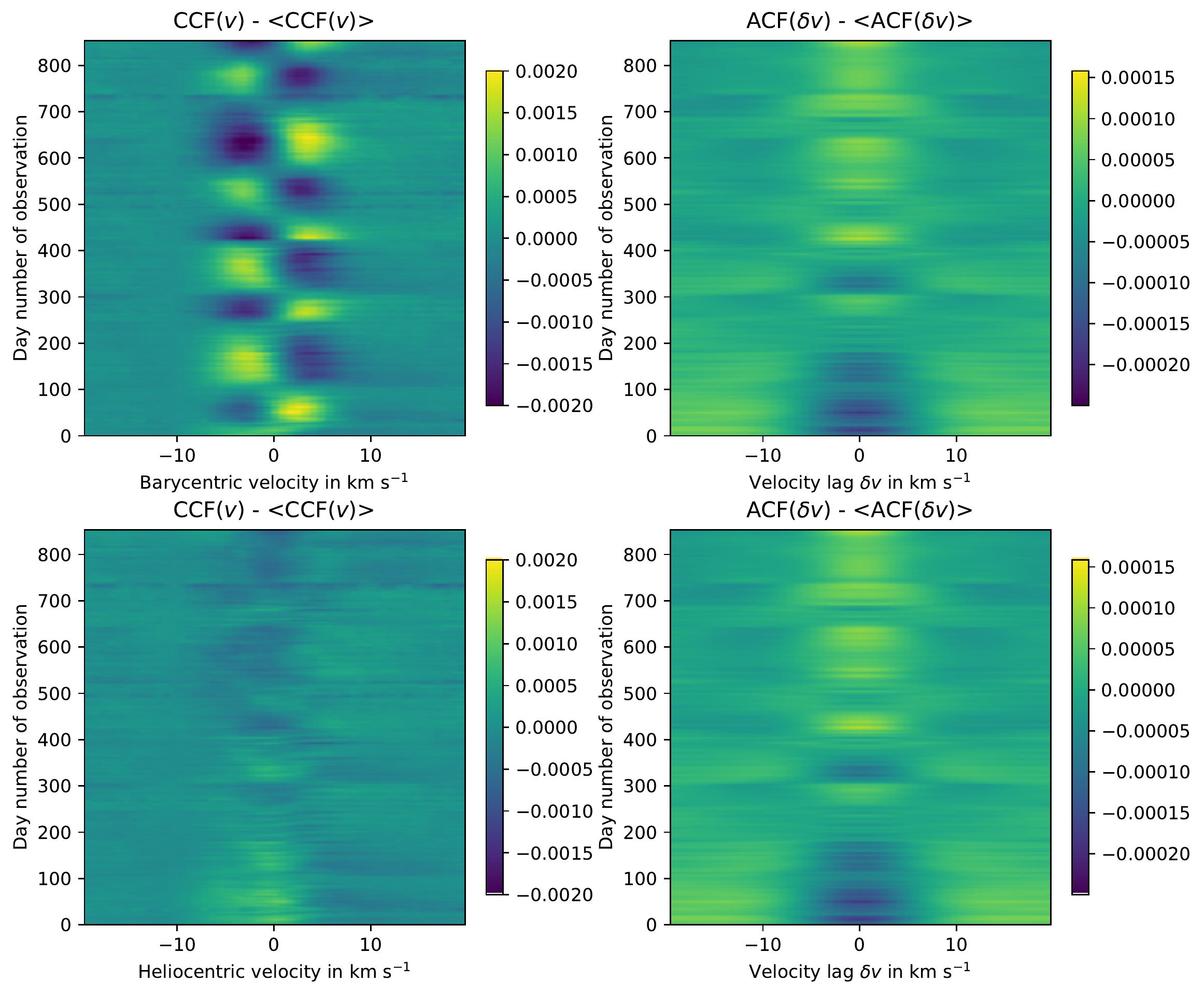}
    \caption{
    Time sequence of residual solar CCFs and the ACFs of the CCFs, spanning the period from 2015 July to 2020 March. 
    Each row represents one day of observation. 
    Days on which no data were obtained are not shown, so although time increases up the vertical axis, the time scale is not linear. The barycentric residual CCF (upper left) shows the solar reflex motion about the solar system barycentre. Dominated by Jupiter, it has a semi-amplitude of 12 m~s$^{-1}$ at Jupiter's synodic period of 398 days. The residual CCF in the heliocentric frame (lower left) shows a secular change in line depth, deepening as the solar activity level declines over the period of observation. The residual ACF (right panels) shows temporal variability that is correlated with that of the CCF (compare left and right panels), but is unchanged by Doppler shifts being applied to the CCF (compare upper right and lower right panels).}
    \label{fig:ccfRes}
\end{figure*}

\section{Cross-correlation function and its autocorrelation}
\label{sec:CCF}

The HARPS, HARPS-N and ESPRESSO Data Reduction Systems (DRS) derive radial velocities from stellar {\'e}chelle spectra by computing the cross-correlation function (CCF) of the spectrum with a digital line mask matched approximately to the spectral type of the target star \citep{1996A&AS..119..373B,2002A&A...388..632P}. 

The data presented in this paper were re-analysed with the \textcolor{black}{ESPRESSO data-reduction pipeline \citep{2021A&A...645A..96P}.} The reduction procedure differs from that used for previous analyses of HARPS-N data \citep[cf. ][]{2019MNRAS.487.1082C} in several important respects. These are described in detail by \citet{2020arXiv200901945D}. The wavelength scale is derived using a new line list tailored for the primary HARPS-N ThAr calibration lamp. A single master wavelength calibration is used for all observations. The wavelength of each pixel in the extracted spectrum is corrected for drift relative to the master calibration using the daily  wavelength calibrations with the primary ThAr lamp, and the simultaneous reference source (either a secondary ThAr lamp or a Fabry-Perot). 
\textcolor{black}{Prior to cross-correlation with a synthetic mask, the pixel wavelengths are transformed to the reference frame of the solar system barycentre along the line of sight to the target. }
The CCFs are estimated on a common grid of pixels at uniformly-spaced intervals of $h=0.82$ km s$^{-1}$ in velocity space, using \textcolor{black}{a blaze-corrected, inverse-variance weighted cross-correlation with a mask of line wavelengths and weights} appropriate to the target spectral type. The CCF sampling interval matches the velocity increment per physical CCD pixel in the instrument. The velocity scale of the CCF is in the reference frame of the solar system barycentre, with the drift correction applied.

Being the cross-correlation of a stellar absorption spectrum with a positive line mask, the CCF computed by the DRS resembles a single stellar absorption line, with a pseudo-continuum level which is \textcolor{black}{accurately and consistently} normalised to unity. 
The resulting CCF profile is fitted with one minus a Gaussian function described by three parameters: central velocity ($v$, relative to the mask), full width at half-maximum depth (FWHM) and central line depth as a fraction of the pseudo-continuum. 
Similarly, the bisector inverse slope (BIS) of the profile \citep{2001A&A...379..279Q} is recorded as a measure of profile asymmetry.  

Stellar activity compromises the fidelity of this method of radial-velocity measurement. The contrast between bright convective upflows in photospheric granules and cooler downflows in intergranular lanes imposes an inherent asymmetry on the line profile \citep{1981A&A....96..345D}. Magnetic activity, the finite lifetime of the granulation pattern and $P$-mode oscillations all cause the already non-Gaussian shape of the observed CCF to vary with time.
As the star rotates, dark starspots produce line-absorption deficits which migrate across the profile from blue to red, introducing time-varying amounts of skew and kurtosis into the spectral-line shapes \citep{1997ApJ...485..319S,2014ApJ...796..132D}. 
These magnetic regions alter the local convective velocity and line-profile asymmetry, as well as the local brightness weighting of the stellar rotation profile \citep{2010A&A...519A..66M}. 
In faculae-dominated stars like the Sun, magnetic suppression of granular convection in faculae causes even stronger time-varying profile asymmetries than sunspots, combining rotational Doppler shifts with foreshortening-dependent changes in the radial-tangential velocity field \citep{2010A&A...519A..66M,2019ApJ...879...55C}. 

Several previous studies 
\citep[e.g.,][]{2012Natur.491..207D,2012MNRAS.419.3147A,2015MNRAS.452.2269R}  
have explored whether the estimated velocities could be improved by decorrelating with other measurements, such as the FWHM or BIS.  
While the FWHM and BIS contain information about the CCF shape, they are not sufficient to describe the detailed changes in the CCF.  
In order to make use of all information contained in the spectra, some studies have suggested analyzing stellar variability by applying a principal-component analysis (PCA) to the observed spectroscopic time series \citep[e.g.,][]{Davis2017,Jones2017}.  
These methods are promising on simulated datasets, but applying this approach to actual observations is challenging due to details of the spectrograph and its calibration process.  
In this paper, we apply the PCA approach to the CCF instead of the raw spectrum. 
This approach leverages extensive investments in developing a robust pipeline to measure the CCF.  
Of course, analyzing only the CCF does reduce the total information content of the spectrum.  
We offer suggestions for how the method could be generalized to extract more information in Section~\ref{sec:discuss}.  

To illustrate the impact of stellar variability on the CCF and to provide a test dataset, we created a sequence of \textcolor{black}{886 daily solar CCFs} from the HARPS-N solar telescope \citep{2014SPIE.9147E..8CC,2016SPIE.9912E..6ZP,2015ApJ...814L..21D}, spanning the period from 2015 July 29 to 2020 March 6.
The resulting distortions of the CCF profile, described in detail by \cite{2019MNRAS.487.1082C}, are seen most clearly in the residuals $R(v_i,t_j)= \ccf - \meanccf$ obtained by subtracting the time-averaged profile of the entire 5-year sequence from each CCF in the sequence \textcolor{black}{(Fig.~\ref{fig:ccfRes}, left-hand panels)}. 

Full details of the HARPS-N solar observations are given by \cite{2019MNRAS.487.1082C}, who used a Gaussian mixture model to assign a probability to each observation that it is unaffected by uneven transparency across the solar disc and corrected the velocities for differential extinction. 
Here we select only those observations with: (1) probabilities greater than 99\% of being good, and (2) with velocity corrections for differential extinction less than 10 cm s$^{-1}$. In  summer these conditions are satisfied for up to 4 hours per day, and in winter for up to 2 hours per day.

\textcolor{black}{Our 5-minute exposure time is dictated by the need to average out solar $p$-mode oscillations. Light from the Sun is gathered by a 76mm objective lens of 200mm focal length, scrambled in an integrating sphere and fed into the spectrograph via an optical-fibre feed to the calibration unit and a neutral-density filter which attenuates the throughput by a further factor 15 \citep{2016SPIE.9912E..6ZP}.
The overall throughput is comparable to night-time HARPS-N exposures for a star of magnitude 5.5, for which we use the same exposure time in good seeing without saturating the detector. This gives SNR$\simeq 350$ or so in the continuum in {\'echelle} order 60, which translates to SNR$\simeq 5000$ in the CCF. For stellar RV observations we use 15-minute blocks of contiguous exposures to mitigate the effects of $p$-mode oscillations. Within the windows that satisfy our selection criteria each day, we select at random a set of three contiguous CCFs to mimic an RV observation of a $V=5.5$ star, and form a weighted average CCF using the square of the mean SNR of the CCF as the weighting factor.} 
Thus, we anticipate that our test data set is dominated by solar magnetic activity, \textcolor{black}{granulation} and/or instrumental issues.  
We encourage future studies to investigate how well the algorithm can mitigate spectral variability on shorter timescales.

\subsection{Translation to the heliocentric frame}
\label{sec:barytohel}

The HARPS-N DRS was designed primarily for stellar radial-velocity measurement, 
\textcolor{black}{computing the CCF in the reference frame of the solar-system barycentre in the direction of the target, as described above.}
As a result, the instantaneous CCF is Doppler-shifted by the component of the Sun's barycentric motion in the observer's direction, as is apparent from the upper-left panel of Fig.~\ref{fig:ccfRes}. To convert CCFs derived from solar spectra to the heliocentric reference frame, the CCF profiles must be shifted by the line-of-sight component $\epsilon$ of the Sun's reflex motion about the barycentre. 

The barycentric to heliocentric velocity corrections were computed using the JPL {\sc horizons} software of \cite{1996DPS....28.2504G}. 

We use a Taylor-series approximation to eliminate the solar barycentric motion from the CCF timeseries, adding scaled derivatives of the instantaneous profile shape at time $t_j$ to the barycentric CCF:
\begin{equation}
\mathbf{C}(v_i+\epsilon,t_j) = \mathbf{C}(v_i,t)+\epsilon\mathbf{ C'}(v_i,t_j)+\frac{\epsilon^2}{2}\mathbf{C''}(v_i,t_j) + \mathcal{O}\epsilon^3.
\label{ccfShift}
\end{equation}

The derivatives are calculated numerically using eqs. \ref{eq:appdfdv} and \ref{eq:appd2fdv2} (see Appendix \ref{sec:appA}). The differences between neighbouring CCF values are substantially less than unity. The barycentric to heliocentric velocity correction is never greater than \textcolor{black}{$\pm 14.7$} m s$^{-1}$, which is much less than the $h=820$ m s$^{-1}$ sampling interval between neighbouring CCF elements. The truncation error in eq. \ref{ccfShift} is therefore significantly less than $\epsilon^3/12h^3\simeq 2.6\times 10^{-8}$.

We validated the fidelity of the shift by calculating heliocentric velocities from the shifted profiles using the methodology of Appendix~\ref{sec:CCFexpansion}. We computed barycentric velocities from the original un-shifted CCFs by the same method, then applied the barycentric to heliocentric velocity correction. The RMS scatter of the difference between the two resulting sets of heliocentric velocities was 0.008~m~s$^{-1}$. The RMS scatter in the difference between the heliocentric velocities calculated from the shifted CCFs and the DRS velocities transformed to the heliocentric frame was \textcolor{black}{0.044}~m~s$^{-1}$.

The resulting CCF timeseries, shown at lower left in Fig.~\ref{fig:ccfRes}, is effectively that of a star with no planets. This image shows that the form of the solar CCF is far from static, with dramatic changes in profile shape taking place on all timescales from days to years. 

\begin{figure}
    \centering
    \includegraphics[width=\columnwidth]{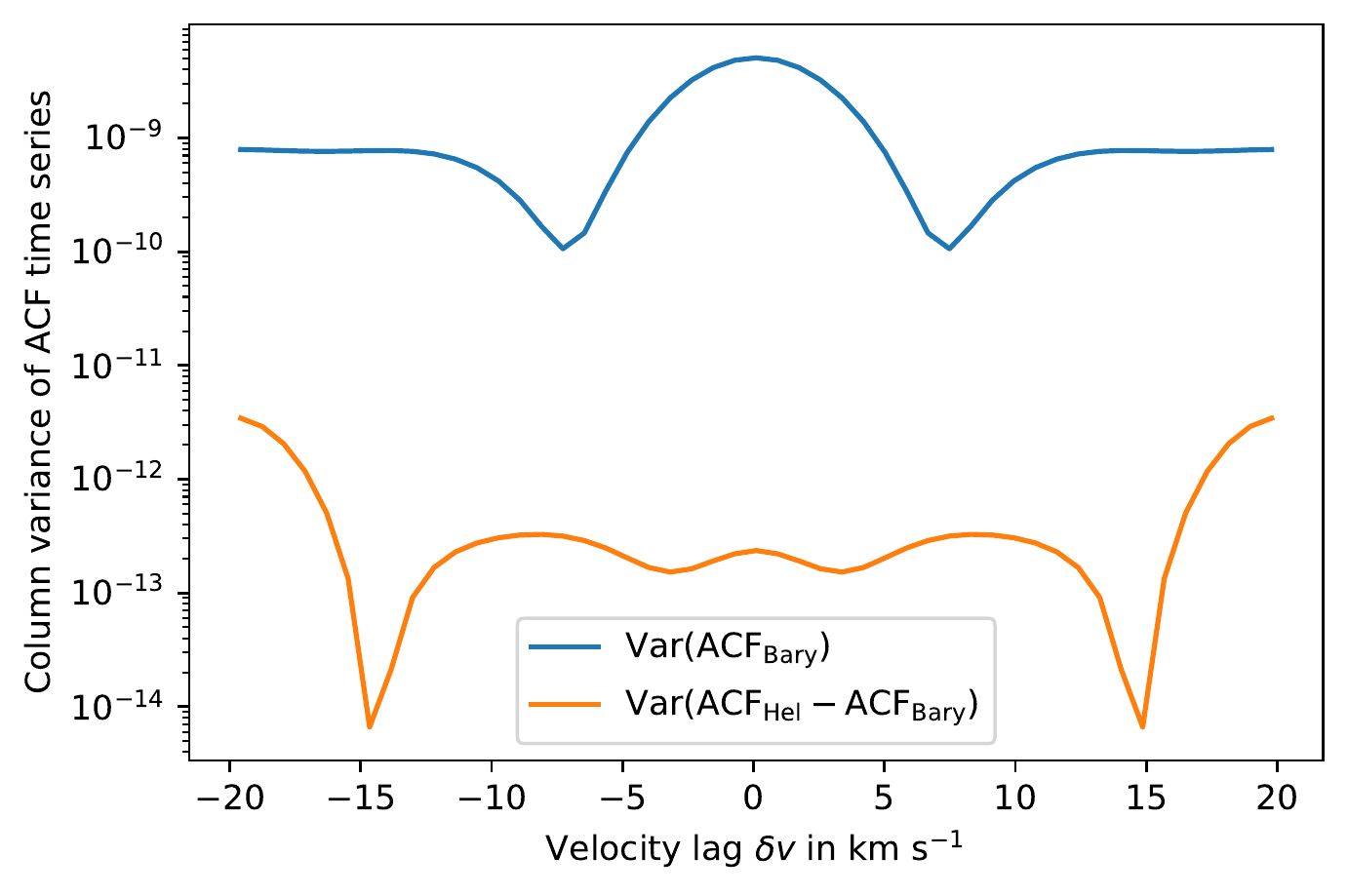}
    \caption{Column variances of the ACF timeseries derived from CCFs in the barycentric frame, compared to column variances of the difference between the heliocentric and barycentric ACFs. The column variance of the difference is 2.5 to 4.5 orders of magnitude smaller than the column variance of either ACF timeseries.}
    \label{fig:acfResDiffVar}
\end{figure}

\begin{figure*}
    \centering
    \includegraphics[width=2\columnwidth]{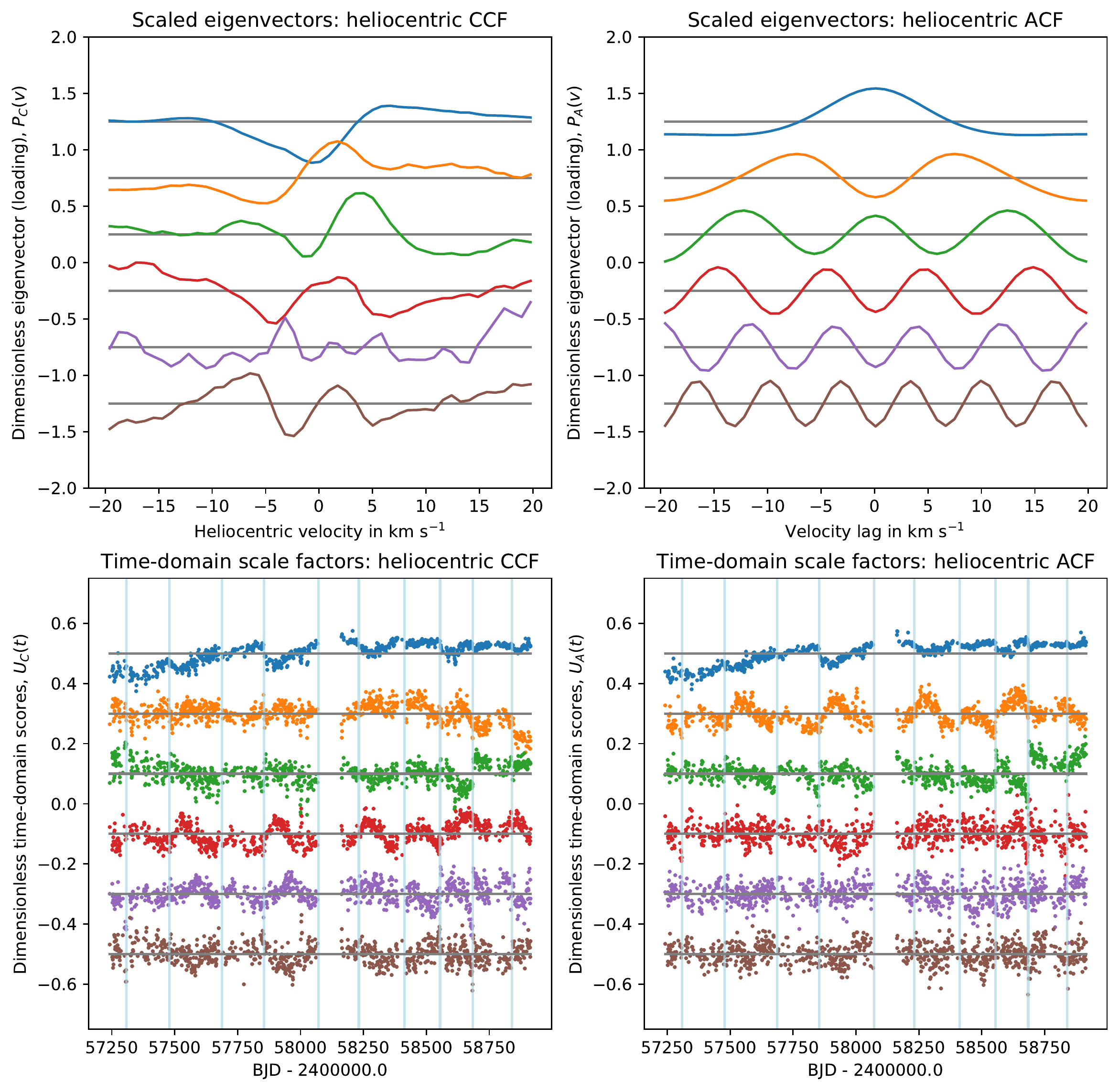}
    \caption{The first 6 basis vectors (loadings) of the singular-value decomposition of the CCF (upper left) and the ACF of the CCFs (upper right) of the heliocentric time series capture the 
    \textcolor{black}{highest-variance}
    stellar and instrumental behaviours. In the lower panels, their \textcolor{black}{time-domain} coefficients (scores) are plotted against barycentric Julian date.  Both the basis vectors and scores are normalised and have been arbitrarily shifted in the vertical direction for clarity. The colours of the scores in the lower panel match the corresponding loadings in the upper panel. The sign of the basis vector (prior to shifting) is arbitrary. Vertical light-blue lines in the lower panels denote the dates of cryostat warm-ups.  
    }
    \label{fig:eigenvectors}
\end{figure*}

\subsection{Autocorrelation of CCF and shift invariance}
\label{sec:ACF}

Since our goal is to separate the effects of genuine dynamical Doppler shifts from spurious shifts caused by line-shape changes, we aim to characterise changes in CCF profile shape in a way that is invariant to translation in velocity space. 
The autocorrelation function (ACF) of the CCF has the desired property that it is invariant to translation \citep{adler1962}.
The ACF $A(\delta v)$ is the expectation value of the vector cross-product of the CCF with itself at a sequence of lags $\delta v$:
\begin{equation}
A(\delta v) = {\rm E}(\mathbf{\mathrm{CCF}}(v) \cdot  \mathbf{\mathrm{CCF}}(v+\delta v))
\label{eq:acf}
\end{equation}
For the sake of brevity, in this manuscript, we refer to the ACF of the CCF as simply ``the ACF''.  

The CCF series has $m$ rows representing individual observations and $l$ columns representing individual velocity bins.
We compute the ACF of every CCF in the time series. This is done by
sequentially shifting the CCF by integer numbers of velocity steps, or CCF ``pixels'', modulo the number $l$ of elements in the CCF, and co-multiplying by the un-shifted CCF:
\begin{equation}
    A(v_{i},t_j) = \sum_{i'=1}^l C(v_{i'},t_j) C(v_{{\rm Mod}(i'-i,l)},t_j).
    \label{eq:acfRoll}
\end{equation}
This \textcolor{black}{set of circular shifts and cross-products} is repeated for every observation, to obtain a time sequence of ACFs. 
The $m$ rows of the ACF time series have the same length $l$ as the original CCFs and are normalised to a mean value of unity. \textcolor{black}{There is sufficient pseudo-continuum to either side of the dip in the CCFs to ensure that this circular autocorrelation procedure is sensitive to long-range correlations while minimising edge effects.}

\textcolor{black}{The right-hand panels of Fig.~\ref{fig:ccfRes} show that, despite the strong differences between the residual CCFs in the barycentric and heliocentric frames, their ACFs are very similar. The similarity is, however, only approximate. The autocorrelation domain is not infinite, and the circular shift method employed is vulnerable to edge effects if the CCFs are strongly shifted. In Fig.~\ref{fig:acfResDiffVar} we compare the column variances of the barycentric ACF timeseries with the column variances of the residuals obtained by subtracting the barycentric from the heliocentric ACF timeseries. We find that the temporal variance of the residual ACF is between 2.5 and 4.5 orders of magnitude smaller than the temporal variance of either the barycentric or heliocentric ACF, at every point in the profile. We conclude that for the purposes of this study, the ACF is {\em effectively} invariant to the solar reflex motion around the solar-system barycentre.}

\subsection{Principal-component analysis of stellar variability}

The principal modes of variability in the CCF can be isolated by calculating the singular-value decomposition (SVD) of the ensemble of CCFs: 
\begin{equation}
\mathbf{C}(v_i,t_j) = \meanccf + \mathbf{U}_C(t_j) \cdot {\rm diag}(\mathbf{S}_C)\cdot \mathbf{P}_C(v_i).
\label{eq:ccfSVD}
\end{equation}
The same method yields the principal components of the ensemble of ACFs:
\begin{equation}
\mathbf{A}(\delta v_i,t_j) = \meanacf + \mathbf{U}_A(t_j) \cdot \rm{diag}(\mathbf{S}_A)\cdot \mathbf{P}_A (\delta \mathit{v_i}).
\label{eq:acfSVD}
\end{equation}
The diagonal matrices $\mathbf{S}_C$ and $\mathbf{S}_A$ list the singular values (eigenvalues) of the principal components in decreasing order. Fig.~\ref{fig:eigenvectors} (top) shows the eigenvectors $\mathbf{P}_{C,k}(v_i)$ and $\mathbf{P}_{A,k}(v_i)$ (also known as loadings) of the leading ($k=1\cdots 6$) principal components of the heliocentric CCF (left) and ACF (right) time series. They represent orthonormal modes of profile variability.

The columns of $\mathbf{U}_{C,k}(t_j)$ and $\mathbf{U}_{A,k}(t_j)$ define an orthonormal basis in the time domain. Each column comprises the coefficients (also known as scores) that define the temporal behaviour of the corresponding eigenvector. 
Fig.~\ref{fig:eigenvectors} (bottom) shows the scores of the leading 6 eigenvectors for all the individual observations in the time-series ensemble, plotted against barycentric Julian date. 
The ACF is calculated in such a way that it is an even function, so its eigenvectors are also even functions. Those of the CCF display a mix of even and odd character. 
Nonetheless, there are strong similarities in the temporal behaviours of their scores.
\\
The scores of the first principal component of both the CCF and the ACF, plotted in blue at the top of all four panels, show a secular \textcolor{black}{upward} trend with a superposed signal of higher frequency. 
The form of the trend, and the shape of the corresponding CCF eigenvector, indicates that this mode of variation affects both the depth and asymmetry of the line profile. It bears a strong resemblance to the variability of the CCF area (i.e. the product of the FWHM and central line depth) noted by \citet{2019MNRAS.487.1082C}. These authors attributed the trend in CCF area to a secular decline in solar network flux and the faster variations to passages of active-region faculae across the solar disc. 
Thus, one sees that the ACF is able to recover a very similar series of scores in a way that is insensitive to line shifts. 
We will exploit this property for separating true Doppler shifts from stellar variability in Section~\ref{sec:scalpels}.  

\begin{figure}
    \centering
    \includegraphics[width=\columnwidth]{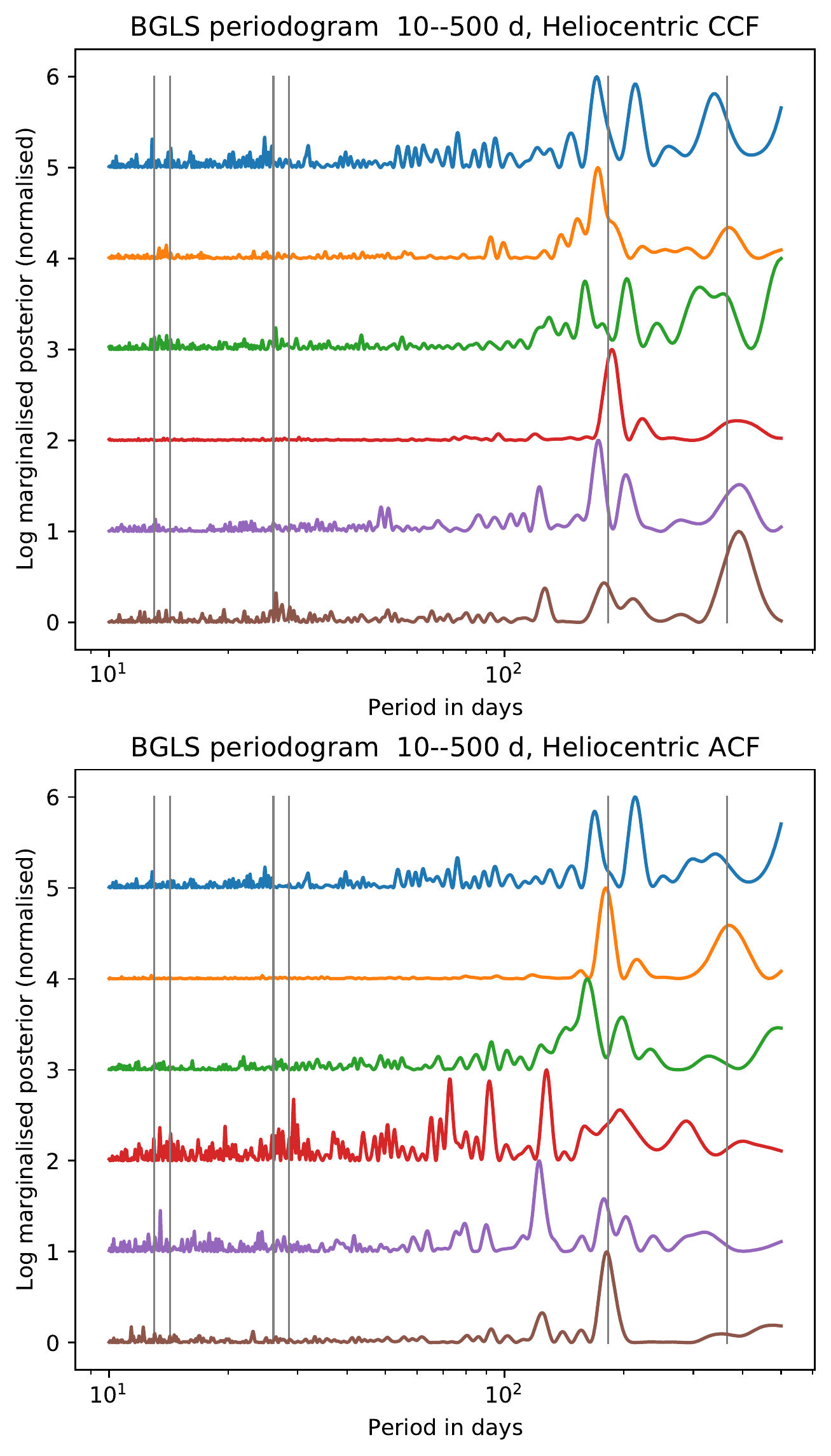}
    \caption{Bayesian Generalised Lomb-Scargle periodograms of the 6 leading principal components of the residual CCF (upper) and ACF (lower) of the heliocentric time series, in the same order as their counterparts in Fig.~\ref{fig:eigenvectors}. The y-axis is the posterior probability density marginalised over the amplitude, phase and zero point of each time series, rescaled to a peak value of unity for display purposes. Several of the principal components of both time series exhibit power around the first harmonic of the solar synodic rotation period, denoted by vertical bars at $P=13.0$ day and $14.25$ day. A similar pair of bars marks the solar rotation period. Other bars show periodicities of 6 months (solar obliquity) and 1 year (Earth orbital eccentricity). Successive traces are offset by 1 unit for clarity.  
    }
\label{fig:ccfBGLS} 
\end{figure}

The time variations of the scores of the second principal component of the ACF (orange traces, second from top in right-hand panels of Fig.~\ref{fig:eigenvectors}) and the \textcolor{black}{fourth principal component of the CCF (red traces, fourth from top in left-hand panels)} are also  similar, though the CCF version appears noisier.  
\cite{2019MNRAS.487.1082C} noted the same pattern of variability in the FWHM of the Gaussian profile fitted to the CCF by the HARPS-N DRS, arising from seasonal changes in the apparent solar rotational broadening. The Earth's orbital eccentricity gives rise to an annual modulation in its orbital angular velocity, and hence the apparent solar rotation rate. The six-month oscillation in the obliquity of the solar rotation axis to the Earth's orbital plane also affects the rotational broadening.
The Bayesian Generalised Lomb-Scargle \citep[][BGLS]{2015A&A...573A.101M} periodogram of the second principal component of the ACF (Fig.~\ref{fig:ccfBGLS}) shows both periods clearly.
The corresponding eigenvector for the CCF resembles the second derivative of the line profile, as expected for the CCF changing in width. 

The third principal component of the ACF (green traces, third from top in right-hand panels of Fig.~\ref{fig:eigenvectors} and the CCF (green, third from top in left-hand panels) also resemble each other. 
They show apparently stochastic discontinuities followed by quasi-exponential decay with a time constant of order a few tens of days. These discontinuities are of instrumental origin. There is a very slow leak in the continuous-flow cryostat of the HARPS-N CCD. The cryostat has to be warmed up approximately twice per year to drive off the water that starts to obstruct the flow of liquid nitrogen. It has been observed that these warm-ups cause a sudden change in the asymmetry of the PSF, which takes a few weeks to decay \citep{2020arXiv200901945D}. During the period of these observations, warm-ups were carried out at JD (24)57161.5, 
57308.5, 
57478.5, 
57687.5, 
57854.5, 
58071.5, 
58231.5, 
58412.5, 
58554.5,
58684.5,
58839.5. 
These clearly coincide with the discontinuities in the third principal component of the ACF.

Examination of the BGLS periodograms (Fig.~\ref{fig:ccfBGLS}) of the leading principal components of the CCF reveals power at half the solar rotation period in \textcolor{black}{the 1st, 2nd, 3rd and 5th principal components. The ACF shows power at this period in the 1st, 4th, and 5th components.} These components probably track profile-shape changes caused by sunspot groups and faculae traversing the visible solar hemisphere. There is surprisingly little power at the solar rotation period in the principal components of either the CCF or the ACF. 

\textcolor{black}{
The CCF shows power at 6 months and/or 1 year in its 2nd and 4th components; the ACF shows power at these periods in the 2nd and 6th components. 
}
Since the heliocentric time series by definition contains no solar reflex motion, these periodic shifts must be also be associated with CCF profile-shape variability arising from Earth's orbital motion. 

The second principal component of the CCF and the third component of the ACF in Fig.~\ref{fig:ccfBGLS}, which we have identified with cryostat warm-ups, shows no power at the solar rotation period or its harmonics in either the CCF or the ACF, but we see significant structure on timescales upwards of 100 days. This indicates that the changes in profile shape caused by cryostat warm-ups are different in character from any form of rotationally-modulated solar activity.

Overall, we see that in the CCF, a simple profile shift would have non-zero projection onto multiple eigenvectors, including those that primarily  represent broadening, skew and kurtosis.  
The eigenvectors of the corresponding components of the CCF have a mix of even and odd characteristics, and their odd parts should therefore affect the measured radial velocities.
However, most of the time variations of the CCF appear broadly similar to those of the shift-invariant profile-shape changes probed by the ACF. 
This raises the possibility that {\em the ACF can be used to deduce the contribution of profile shape changes to the measured radial velocities}. 
We conclude that, at least for the heliocentric solar time series, principal-component analysis of the ACF could provide an effective means of separating the effects of dynamical shift from those of stellar and instrumental profile variability. 

\section{The radial-velocity response to ACF time-domain variations}
\label{sec:scalpels}

To achieve this, we treat the time-series of scores for each principal component of the ACF as the coefficients of a set of unknown eigenvectors representing orthogonal modes of variability in the shape of the CCF. These unknown eigenvectors of the CCF will affect the measured radial velocity to a greater or lesser degree depending on whether they are predominantly of even or odd character.

\textcolor{black}{
\subsection{Projection into the ACF time-domain subspace}
\label{sec:projection}
}

The set of radial-velocity observations has $m$ elements, and can be thought of as a vector $\mathbf{v}_{\rm obs}$ belonging to an $m$-dimensional space $S$. The $l$ orthonormal columns of $\mathbf{U}_A$ have the same dimension as $S$, and define an $l$-dimensional subspace $U\subset S$, centred on the origin of $S$.

We first subtract the inverse-variance weighted mean $\left<v\right>_{\rm obs}$ from the vector $\mathbf{v}_{\rm obs}$ of radial velocities measured with the data-reduction pipeline, to ensure orthogonality.
We project the difference $\mathbf{v}_{\rm obs}-\left<v\right>_{\rm obs}$ on to the time-domain manifold spanned by the basis formed by the columns of the matrix $\mathbf{U}_A$. \textcolor{black}{Each row of $\mathbf{U}_A$ corresponds to an individual observation identified with a unique time-stamp. For conciseness we refer to this observation-identifier space as the "time domain".}
The inner product of the $k$th column of $\mathbf{U}_A$ (i.e., the scores associated with $k$th basis vector for ACF decomposition) with the velocities yields the response $\mathbf{\alpha}_{k}=\mathbf{U}^T_{A,k}\cdot(\mathbf{v}_{\rm obs}-\left<v\right>_{\rm obs})$ of the radial velocity to the time variation of $\mathbf{U}_{A,k}$. The vector of response factors is then
\begin{equation}
\hat{\mathbf{\alpha}} = \mathbf{U}_A^T\cdot(\mathbf{v}_{\rm obs}-\left<v\right>_{\rm obs}).
\label{eq:alpha}
\end{equation}

The sum of the scaled velocity contributions from all principal components of the ACF is then $\mathbf{v}_\parallel = \mathbf{U}_A\cdot\hat{\mathbf{\alpha}}$. This velocity vector $\mathbf{v}_\parallel$ lies within the subspace $U$, and gives a complete model of the radial-velocity perturbations arising from the changes in profile shape to which the ACF is sensitive:
\begin{equation}
    \mathbf{v}_\parallel=\mathbf{U}_A\cdot
    \mathbf{U}_A^T\cdot(\mathbf{v}_{\rm obs}-\left<v\right>_{\rm obs}).
    \label{eq:acfRVproj}
\end{equation}
\textcolor{black}{The product $P_\parallel = \mathbf{U}_A\cdot\mathbf{U}_A^T$ is a projection operator. The operator $P_\perp = (\mathbf{I}- \mathbf{U}_A\cdot\mathbf{U}_A^T)$ projects onto the subspace orthogonal to $\mathbf{U}_A$.
The residual velocities
\begin{equation}
    \mathbf{v}_\perp = P_\perp\cdot(\mathbf{v}_{\rm obs}-\left<v\right>_{\rm obs}) =\mathbf{v}_{\rm obs}-\left<v\right>_{\rm obs} - \mathbf{v}_\parallel
    \label{eq:perpproj}
\end{equation}
lie} outside the subspace $\mathbf{U}_A$, and therefore $\mathbf{v}_\perp$ preserves the shifts to which the ACF is insensitive. \textcolor{black}{Information is, however, lost in this process. The resulting velocities $\mathbf{v}_\perp$ are biased down by the projection of the velocities onto $\mathbf{U}_A$, as discussed in Section~\ref{sec:injtests}.}

\begin{figure}
    \centering
    \includegraphics[width=\columnwidth]{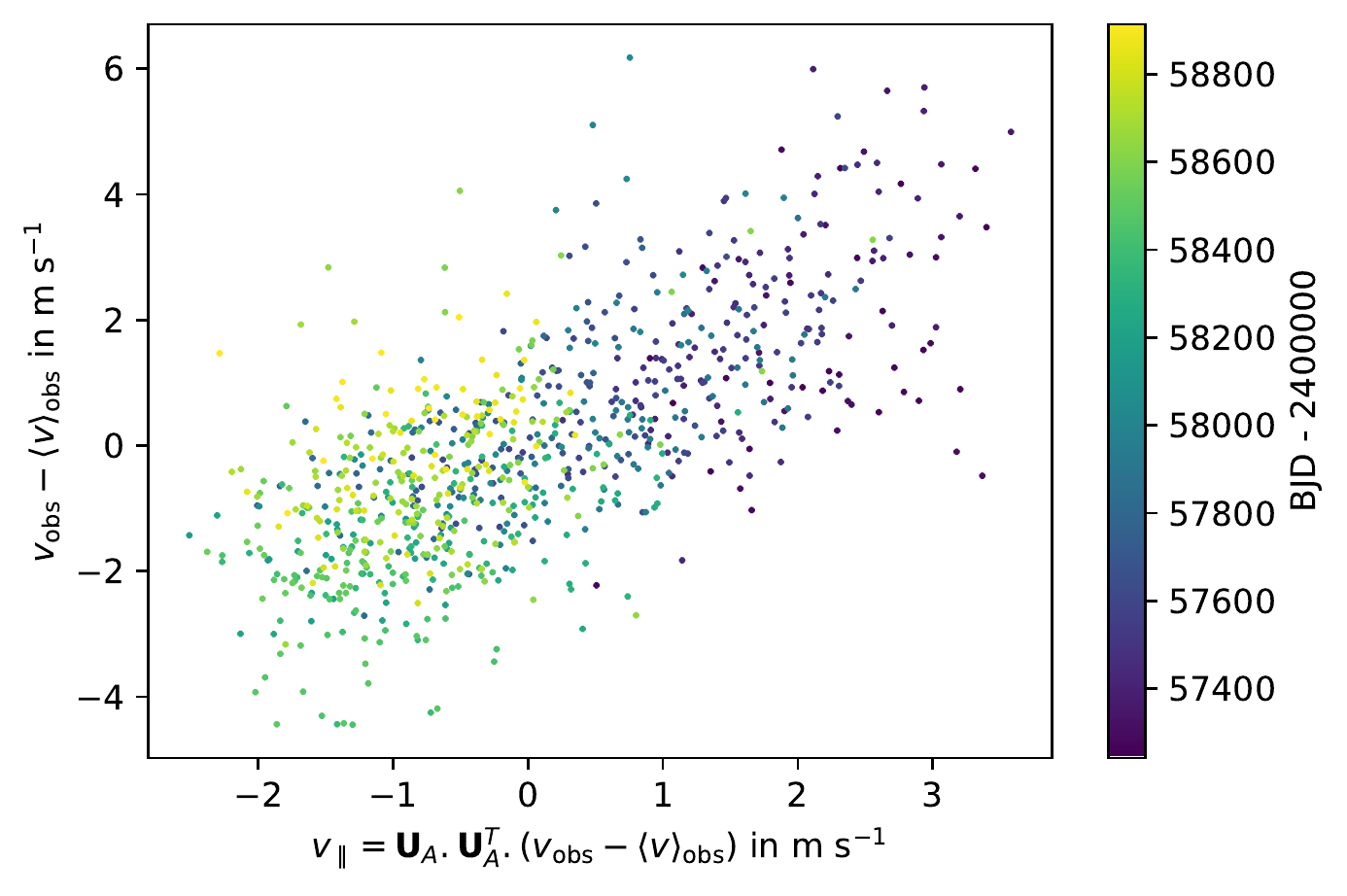}
    \caption{Observed velocities $\mathbf{v}_{\rm obs}$ plotted against the shape-driven velocity component $\mathbf{v}_\parallel$ computed using {\sc scalpels} projection. The individual points are colour-coded by date of observation.   
    }
    \label{fig:rv_correlation}
\end{figure}

\textcolor{black}{From here on, we refer to $\mathbf{v}_{\rm obs}$ as the ``measured'' or ``observed'' velocities"; $\mathbf{v}_\parallel$ as the ``model'' or ``shape-driven'' velocities; and $\mathbf{v}_\perp$ as the ``shift-driven'' velocities. } Fig.~\ref{fig:rv_correlation} shows that the shape-driven velocity perturbations, $\mathbf{v}_\parallel$, are strongly correlated with the observed velocities, reproducing faithfully the long-term and short-term fluctuations dominated by stellar activity. 

Given a set of measured radial velocities ($\mathbf{v}_{\rm obs}$) and the corresponding array of CCFs from which they were derived, \textcolor{black}{equations \ref{eq:acfRoll}, \ref{eq:acfSVD}, \ref{eq:acfRVproj} and \ref{eq:perpproj}} constitute a simple \textcolor{black}{linear projection method} for deriving the shape-driven perturbations to the radial velocity ($\mathbf{v}_\parallel$), so as to provide a substantially cleaned set of shift-driven radial velocities ($\mathbf{v}_\perp$). 
\vspace{1cm}
\subsection{Outlier clipping}
\label{sec:outclip}

The shape of the CCF is sensitive to more than just solar activity. Changes in spectrograph focus can affect the FWHM of the CCF, while cryostat warm-ups perturb the skewness of the profile. Noisy CCFs, saturated exposures, or undetected cloud obscuration of part of the solar disc, can also cause temporary profile distortions which may not correlate with any of the 
\textcolor{black}{highest-variance}
principal components. 

Such anomalous observations may indeed generate their own basis functions when SVD is applied to the ACF time-series. Their coefficients are normally close to zero, except when an anomaly occurs. They then appear as outliers in the corresponding columns of $\mathbf{U}_A$. Such points can be masked as bad (0) if their absolute deviations lie further from the median value of the column than a specified number of median absolute deviations (MAD), and good (1) otherwise. If even one of the coefficients for an observation is an extreme outlier, it is likely that the entire observation is contaminated. We therefore create a one-dimensional rejection mask in the time domain from the product of the column masks. For the solar data we found that clipping at 6 times the MAD within each column of $\mathbf{U}_A$ provided a stable set of basis vectors at the cost of reducing the total number of usable days of observation from \textcolor{black}{886 to 853}. 

This clipping procedure ensures a clean set of basis vectors, but does not detect outliers caused by unwanted velocity shifts, such as might be caused by an anomalous drift measurement. If present, these must be identified and clipped separately.

\textcolor{black}{
\subsection{Rank reduction and column re-ordering}
\label{sec:rankreduce}
}

Following outlier clipping and masking, the singular-value decomposition of both the CCF and the ACF is re-computed from the surviving observations. It should be noted that all figures in this paper from Fig. 1 onward are based on the masked dataset only.

The subspace defined by $\mathbf{U}_A$ has as many dimensions as there are pixels in each row of the ACF array. \textcolor{black}{The CCF and the ACF, however, only display a small number of modes of variability that are detectable above the noise level. Only the highest-variance components of $\mathbf{U}_A$ are needed to capture adequately the shape changes in the ACF. The remaining low-variance components serve only to fit noise. These low-variance components may show spurious correlations with the RV signal, leading to over-fitting of $\mathbf{v}_\parallel$.} It is therefore possible (and desirable) to model the activity signal adequately and avoid over-fitting noise using a reduced number of dimensions.

\textcolor{black}{To determine the optimal size of the null space, we used leave-one-out cross-validation \citep{2014LOOCV}.
Holding out each row $\mathbf{A}_j$ of the ACF in turn, we decompose the remaining rows and compute the singular-value decomposition:
\begin{equation}
    \mathbf{A}_{i\ne j}= \mathbf{U}_{i\ne j}\cdot{\rm Diag}({S}_{i\ne j})\cdot{V}_{i\ne j}.
    \label{eq:a_loocv}
\end{equation}
We reconstruct an estimate $\hat{\mathbf{U}}_j$ of the missing $j$th row of $\mathbf{U}_A$ by fitting the eigenvectors and eigenvalues to the $j$th row of the ACF: 
\begin{equation}
    \hat{\mathbf{U}}_j = \frac{(\mathbf{A}_j\cdot{V}^T_{i\ne j})}{({\rm Diag}({S}_{i\ne j})\cdot{V}_{i\ne j})\cdot{V}^T_{i\ne j}}.
    \label{eq:u_loocv}
\end{equation}
}

\begin{figure}
    \centering
    \includegraphics[width=\columnwidth]{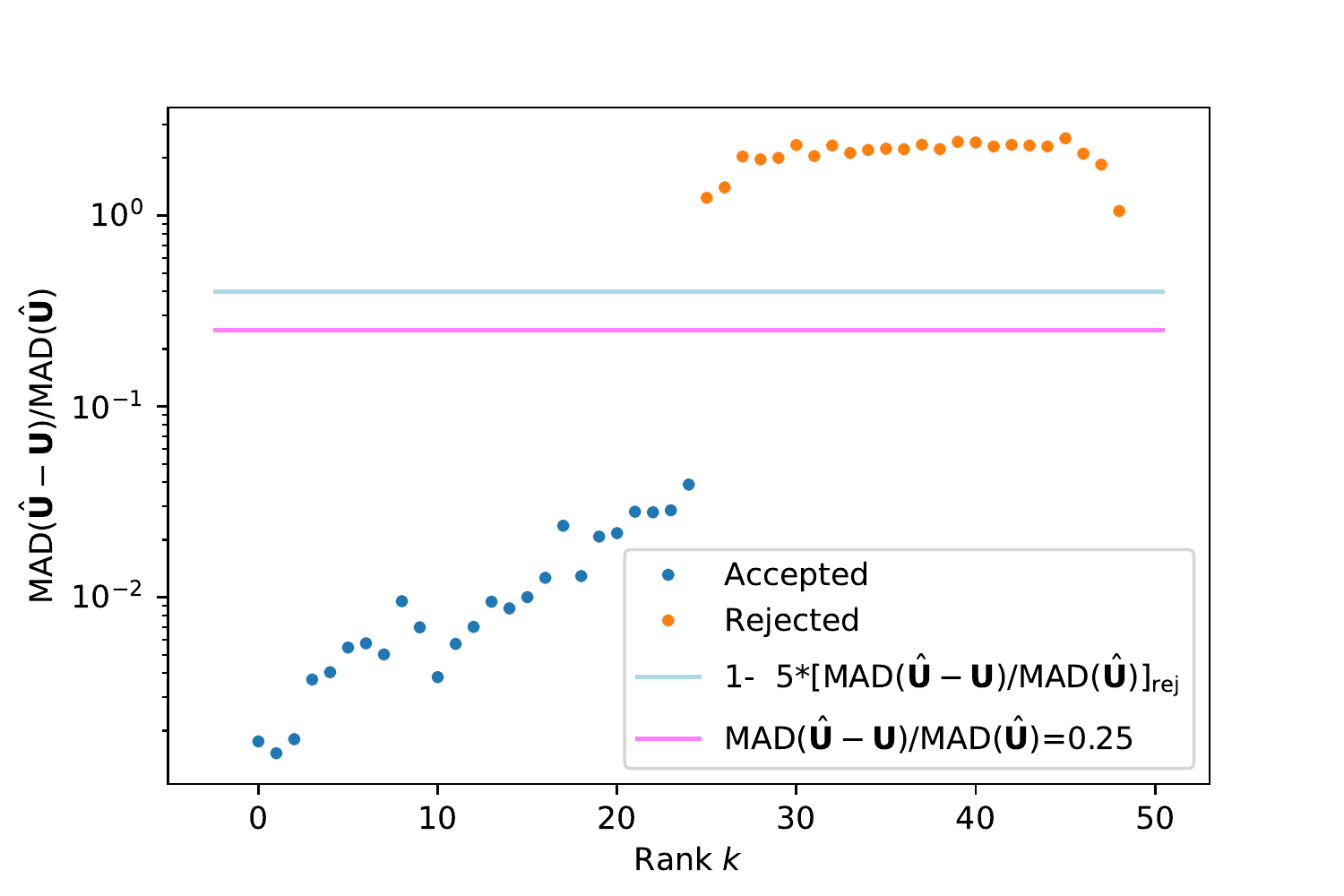}
    \caption{\textcolor{black}{Rank reduction with leave-one-out cross-validation. The ratio of the
median absolute deviation (MAD) of $\mathbf{U}^T_k-\hat{\mathbf{U}}^T_{A,k}$ to ${\rm MAD}(\hat{\mathbf{U}}^T_k)$ rises sharply to values above unity for values of $k>k_{\rm crit}$. Here $k_{\rm crit}=25$.}}
    \label{fig:LOOCV}
\end{figure}

\textcolor{black}{After having repeated this procedure for all rows, we find that the reconstruction of the $k$th column $\hat{\mathbf{U}}^T_k$ reproduces $\mathbf{U}^T_{A,k}$ with good fidelity for $k<25$ or so. The ratio of the
median absolute deviation (MAD) of $\mathbf{U}^T_k-\hat{\mathbf{U}}^T_k$ to ${\rm MAD}(\hat{\mathbf{U}}^T_k)$ rises to values close to unity for values of $k>k_{\rm crit}$ for which the leave-one-out cross-validation indicates that the reconstruction is poor, as shown in Fig.~\ref{fig:LOOCV}.}

\textcolor{black}{Following projection of the RV data into the reduced space defined by the surviving columns of $\mathbf{\hat{U}}$, we find that the quality of the fit between the RV data and the shape model $\mathbf{v}_\parallel$ improves rapidly at first, reaches a minimum then increases gradually as more velocity components are added to the model and overfitting starts to degrade the solution. In other words, we need even fewer principal components to model $\mathbf{v}_\parallel$ than we need to reproduce the ACF itself.}

\begin{figure}
    \centering
    \includegraphics[width=\columnwidth]{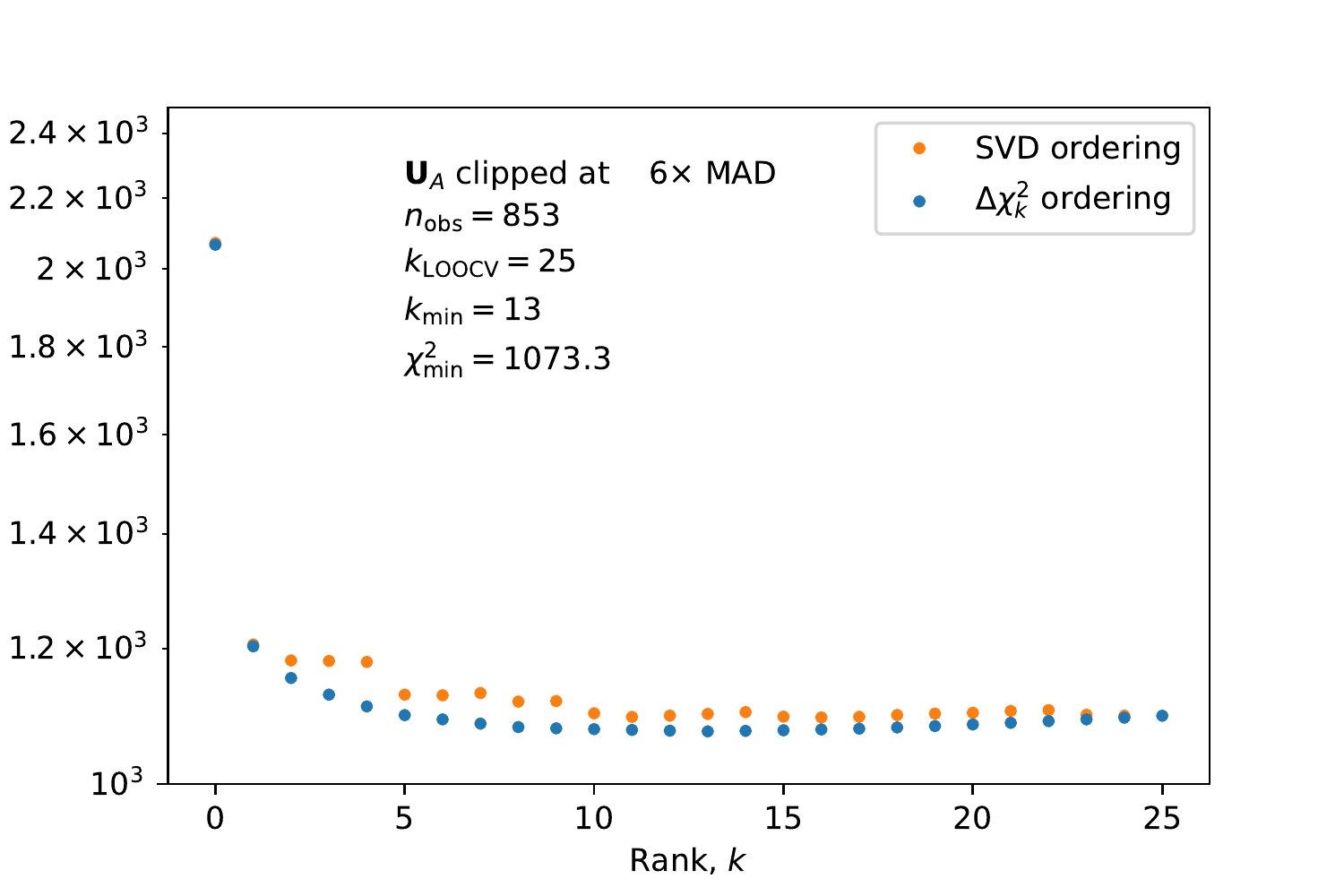}
    \caption{\textcolor{black}{As more dimensions are added to the subspace defined by the time-domain coefficients $\hat{\mathbf{U}}$ of the principal components of the ACF, the $\chi^2$ of $\mathbf{v}_\perp$ decreases to a minimum value, then increases gradually as overfitting degrades the solution. The minimum is reached more rapidly when the columns of U are sorted in order of decreasing $\delta\chi^2$ (blue) rather than in order of their corresponding singular values (orange). With $\delta\chi^2$ sorting, the optimal size of the null space is defined by the minimum at $k=13$.}}
    \label{fig:rankchisq}
\end{figure}

\textcolor{black}{SVD orders the principal components of the ACF in descending order of their eigenvalues $\mathbf{S}_A$. 
This ordering does not take the velocity projection into account, so the ordering of principal components does not reflect accurately their contributions to the radial velocity. 
Instead, we reorder the columns of $\mathbf{\hat{U}}$ into the sequence that gives the fastest decrease in $\chi^2$, obtaining the optimal fit to the RV data with the smallest number of basis vectors, as shown in Fig.~\ref{fig:rankchisq}. }

\begin{figure*}
    \centering
    \includegraphics[width=2\columnwidth]{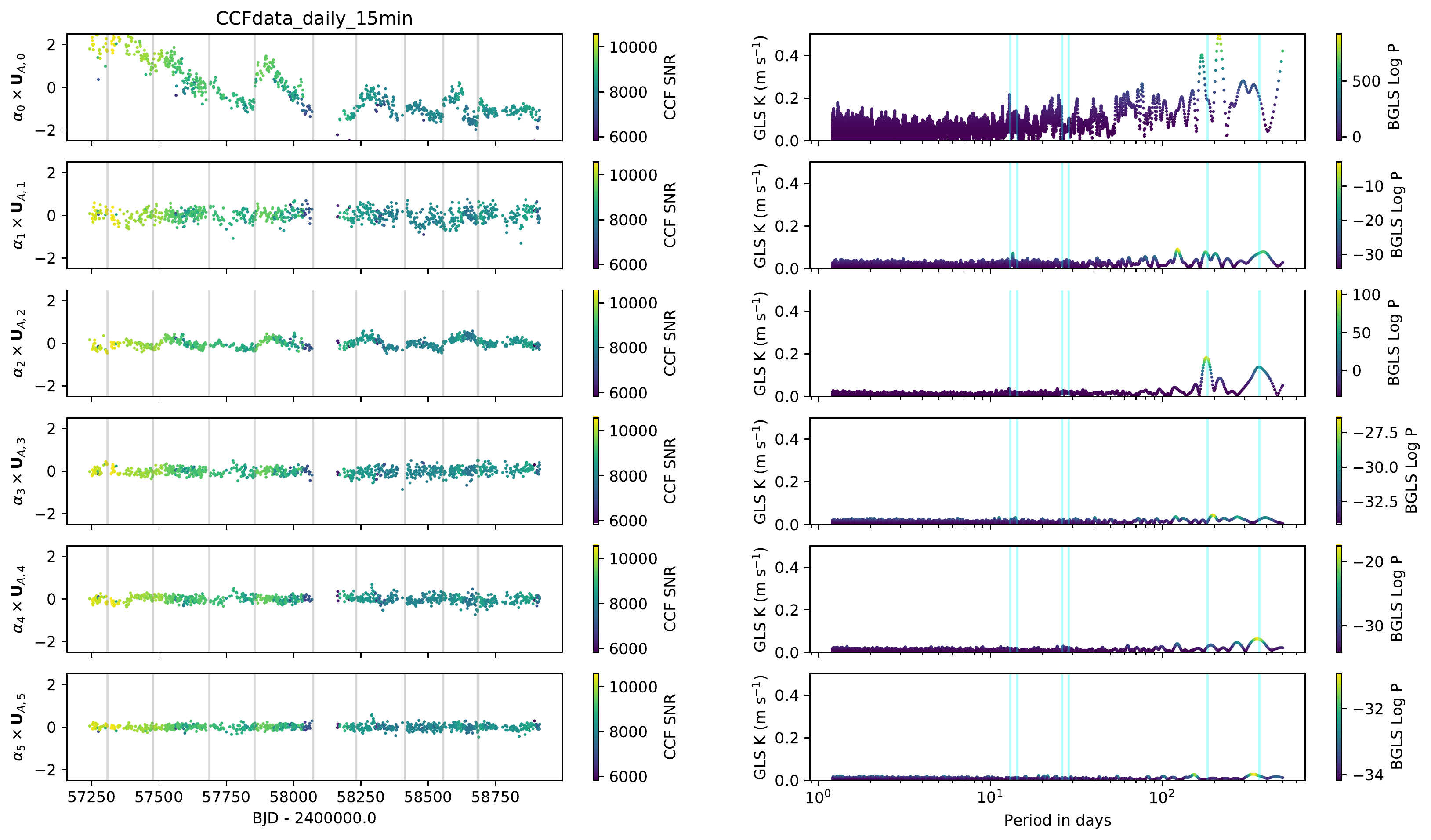}
    \caption{\textcolor{black}{
    Left panel: The first 6 vector components of the projection of the solar heliocentric radial velocities (in m~s${-1}$) into the rank-reduced ACF time-domain subspace, ordered by their projection coefficients $\hat{\alpha}$ and colour-coded by the SNR of the CCF. Vertical grey lines denote the dates of cryostat warm-ups. Right panel: Generalised Lomb-Scargle $K$ amplitude periodograms of the same six time series, colour-coded by Bayesian GLS likelihood \citep{2015A&A...573A.101M}. The vertical blue lines are at periods of 13.0, 14.25, 26.0, 28.5, 182.63 and 365.25 days. Power is seen near the solar rotation period in the first component. Weak power is seen at half the rotation period in the second component. The third component shows the annual and six-monthly modulation of the CCF FWHM caused by Earth's orbital eccentricity and the solar obliquity respectively. Annual signals are also present in the 2nd and 5th components.}
    }
    \label{fig:rvproj}
\end{figure*}

\citet{Davis2017} found that 4 or 5 principal components were sufficient to capture  the temporal behaviour of synthetic spectra produced by a noise-free SOAP2.0 \citep{2014ApJ...796..132D} simulation of starspot activity and facular suppression of convective blueshift on a rotating stellar model. 
As we have seen, the HARPS-N solar data contain additional profile distortions arising from changes in the instrument and Earth's orbital motion, so more principal components are needed. 

We find a good compromise between outlier clipping, number of surviving days of observation, and minimal number of basis vectors when we clip $\mathbf{U}_A$ at 6 times the MAD, as noted above. 
\textcolor{black}{With $\delta\chi^2$ reordering, the $\chi^2$ of $\mathbf{v}_\perp$is minimised at $k_{\rm min}=13$. 
We therefore use the 13 leading principal components of $\mathbf{\hat{U}}_A$, ordered by $\delta\chi^2$, to define the null space. We note, however, that the results that follow are only very weakly sensitive to the number of principal components used over the range $6 < k < 13$ or so.}

The projections of the observed velocities on to the components of $\mathbf{U}_A$ corresponding to the 6 largest values of $\hat{\mathbf{\alpha}}$ are plotted against time in Fig.~\ref{fig:rvproj}.  

\begin{figure*}
    \centering
    \includegraphics[width=\columnwidth]{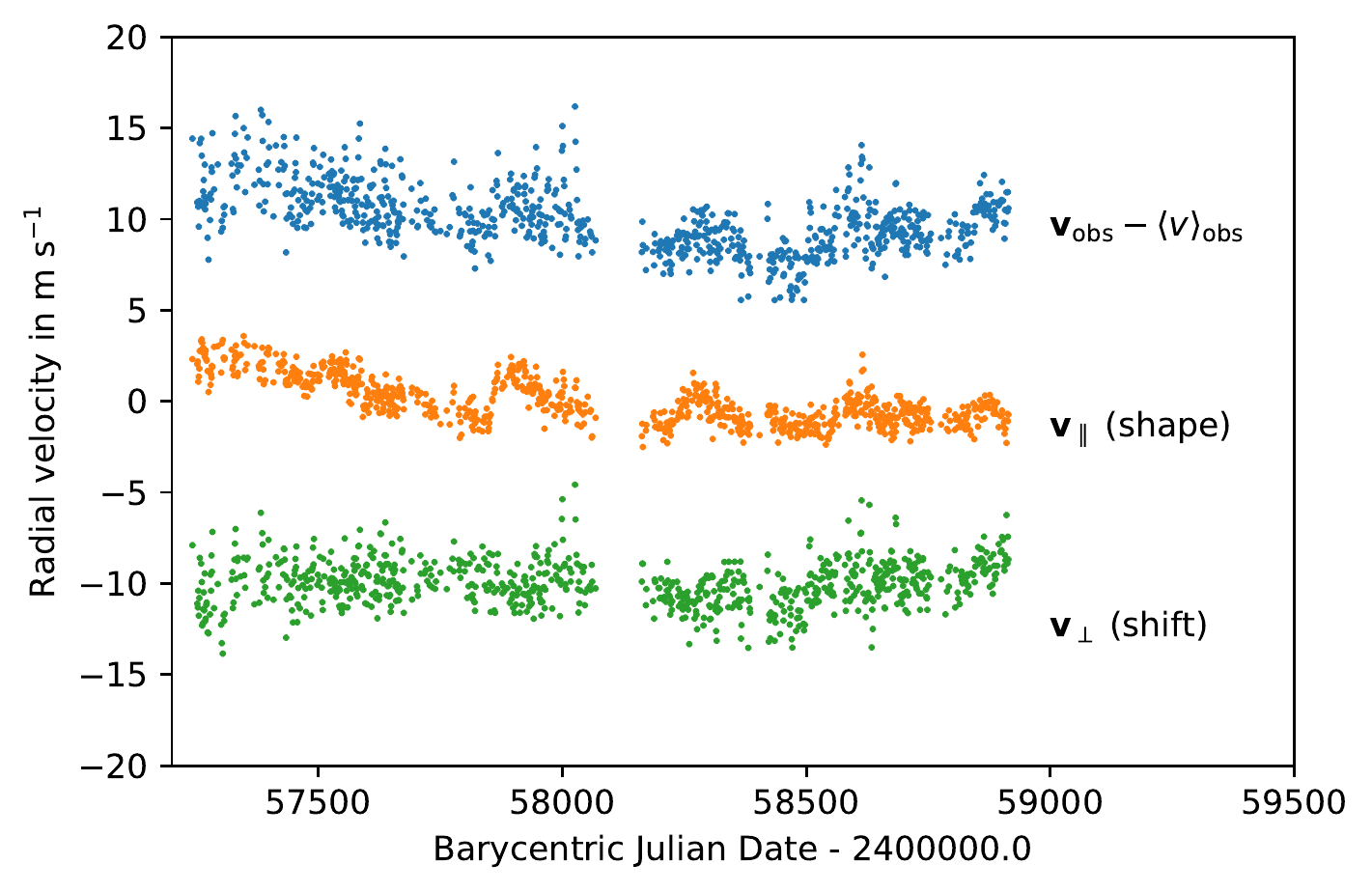}
    \includegraphics[width=\columnwidth]{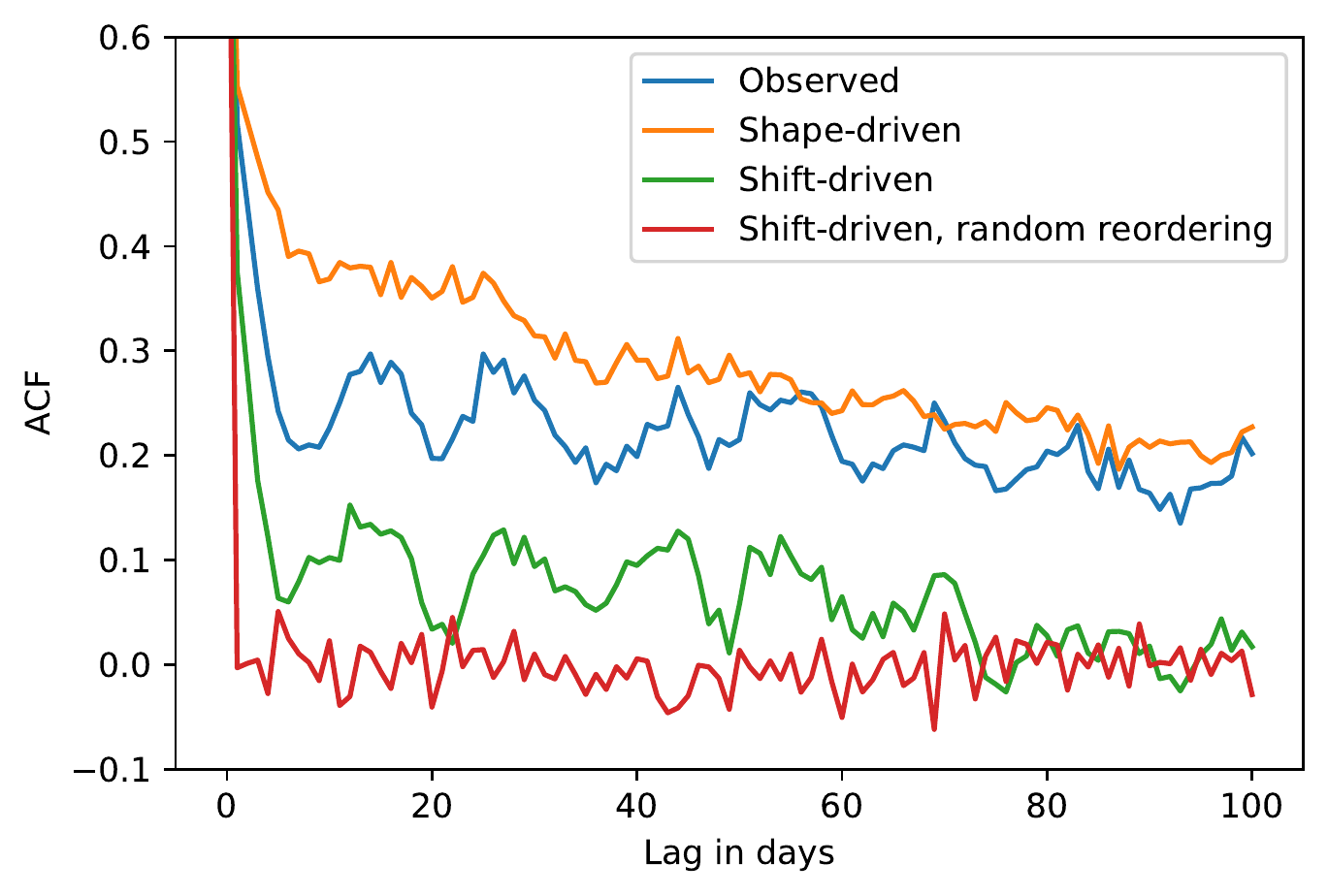}
    \includegraphics[width=\columnwidth]{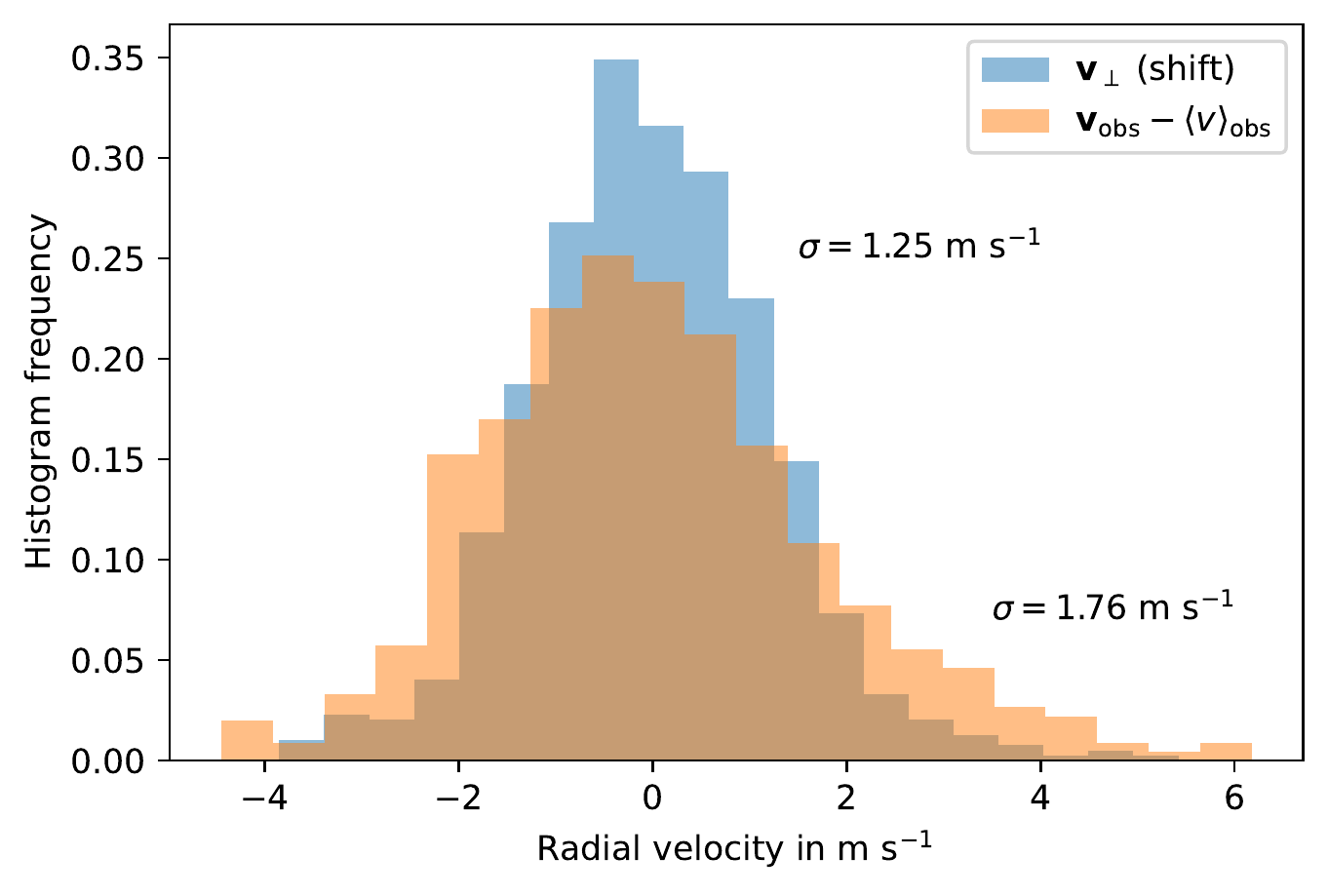}
    \includegraphics[width=\columnwidth]{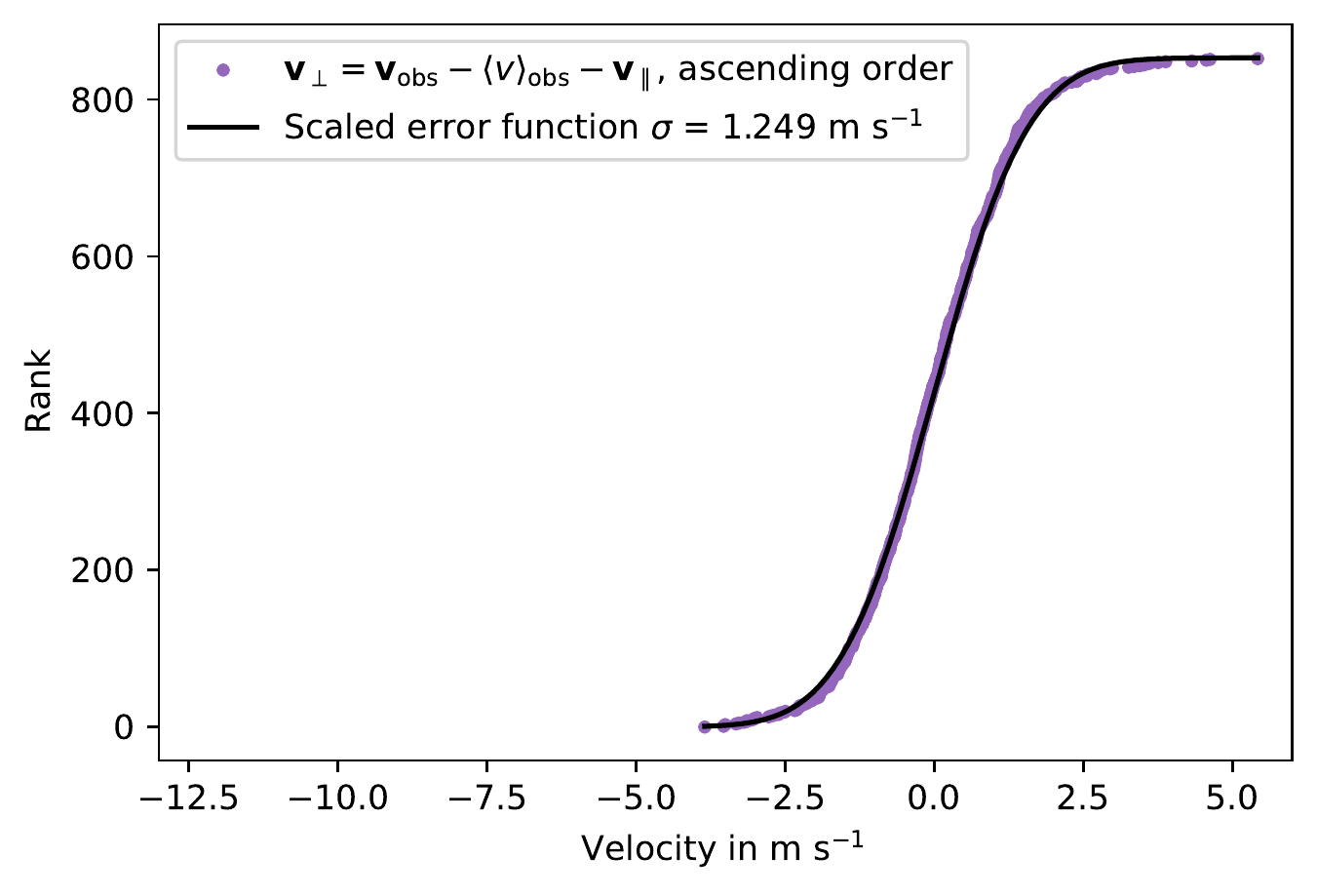}
    \caption{
    Upper left: observed radial velocities transformed to the heliocentric reference frame, together with their {\sc scalpels}-separated shape-driven and shift-driven velocity components, offset by $\pm 10$~m~s$^{-1}$ for clarity. 
    \textcolor{black}{Upper right: Autocorrelation functions of the radial velocities, corrected for missing dates of observation and normalised to unity at zero lag. The original radial velocities (blue) are seen to be strongly correlated at all time lags, and the shape-driven velocities (orange) even more so.  
    The autocorrelation of the shift-driven velocities decays substantially more rapidly than the observed or shape-driven velocities. Any correlation at lags longer than $\sim70$ days is negligible.
    The red curve is the ACF of the shift-driven velocities shuffled into random order, and is effectively uncorrelated.}
    Lower left: Histograms show that the RMS scatter has been reduced from \textcolor{black}{1.76~m~s$^{-1}$ in the original data to 1.25~m~s$^{-1}$ in the shift-driven velocities. 
    Lower right: shift-driven velocities sorted in ascending order and overplotted with the cumulative normal distribution with $\sigma=1.25$~m~s$^{-1}$. }
    }
    \label{fig:velHel}
\end{figure*}

\begin{figure}
    \centering
    \includegraphics[width=\columnwidth]{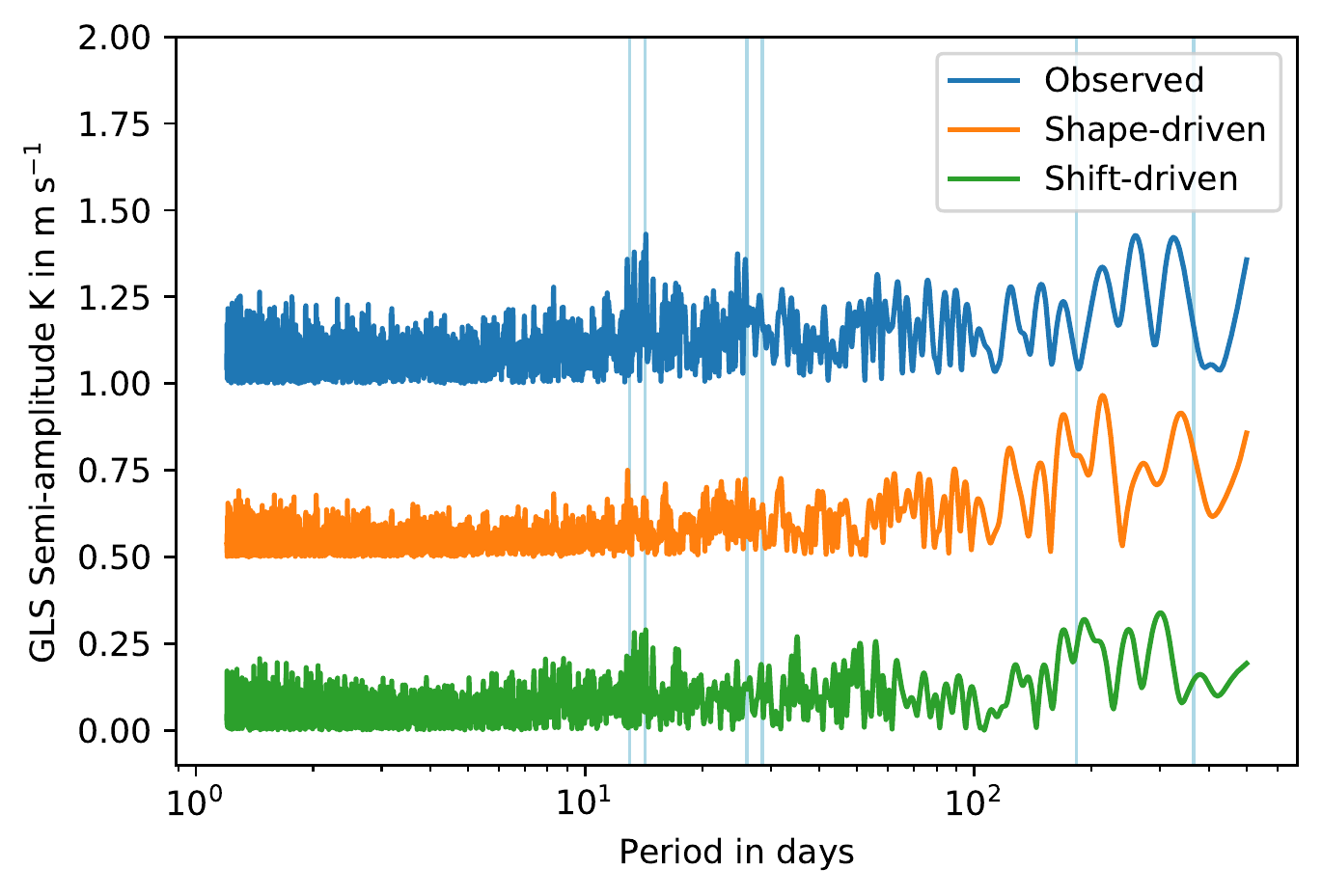}
    \caption{Semi-amplitude periodogram demonstrating the value of signal separation by projection of the observed radial velocities on to the principal time-domain components of the ACF. The top periodogram (blue) is for the measured velocities derived from the CCF. 
    The second trace (orange) is for the shape-driven velocities $\mathbf{v}_\parallel$ produced by the {\sc scalpels} projection.  
    The third periodogram (green) is that of the shift-driven velocities $\mathbf{v}_\perp$ remaining after subtraction of the shape-driven velocities from the observations. \textcolor{black}{Light-blue bars denote the approximate ranges of the solar rotation period and its first harmonic, and periods of 6 months and 1 year.}}
    \label{fig:periodogramHel}
\end{figure}
\vspace{1cm}
\subsection{{\sc scalpels} analysis of solar data}
\label{sec:scalpels_solar}

\textcolor{black}{We refer to the projection of the observed velocities on to the orthogonal complement of the time-domain scores $\mathbf{U}_A$ of the ACF together with the outlier clipping (Sec.~\ref{sec:outclip}) and rank reduction (Sec.~\ref{sec:rankreduce}) algorithms collectively as {\em Self-Correlation Analysis of Line Profiles for Extracting Low-amplitude Shifts} ({\sc scalpels}). The reader is referred to Appendix~\ref{sec:appblind} for a concise listing of the main steps of the algorithm.}

In a blind radial-velocity survey, planet-candidate detection is typically conducted using analysis of some form of periodogram such as Lomb-Scargle \citep[e.g.][]{1976Ap&SS..39..447L, 1982ApJ...263..835S, 2009A&A...496..577Z} or marginalized posterior versus orbital period \citep[e.g.][]{2015A&A...573A.101M} computed from the radial velocities.  Periodogram peaks are identified and fitted with Keplerian orbit models. This method is, however, susceptible to confusion with rotationally-modulated signals from the host star.

To assess
the effectiveness of the {\sc scalpels} algorithm for suppressing stellar noise, we apply it to the daily-averaged solar radial velocities in the heliocentric frame. In the absence of any dynamical shifts or instrumental calibration drift, the measured radial velocities should show only variations caused by line-profile shape changes arising from solar activity, changes in the instrumental point-spread function, and changes in the apparent solar rotation rate arising from Earth's orbital eccentricity and the solar obliquity. Given that the heliocentric solar data set contains no planet signals, a frequency search for candidate periodic signals provides a means of establishing the planet-detection threshold for comparable datasets.

In Fig.~\ref{fig:velHel} we show the observed heliocentric radial velocities minus their own mean, together with the shape-driven velocities obtained by {\sc scalpels} projection and the shift-driven velocity difference between the two. 
The histograms of the observed velocities and the shift-driven velocities are also shown. The distribution of observed velocities is severely non-Gaussian, with a bimodal character arising from short-term (days-weeks) and long-term activity (years) variability.

After subtracting the {\sc scalpels}-identified shape-driven velocity residual velocities $\mathbf{v}_\parallel$, 
the shift-driven velocities $\mathbf{v}_\perp$ are nearly constant with respect to time, with a local RMS scatter of \textcolor{black}{$1.25$~m~s$^{-1}$}. 
The Anderson-Darling 1-sample test \citep{1987JASA...82...918} indicates that the distribution of the shift-driven velocities
is indistinguishable from a normal distribution (Fig.~\ref{fig:velHel}, lower panels), with standard deviation \textcolor{black}{$\sigma=1.25$~m~s$^{-1}$}.  

Any stellar activity signature remaining in the shift-driven velocity time series is likely to show temporal correlations and departures from uncorrelated Gaussian noise. There appear to be  weakly correlated residuals with amplitudes of a few tens of cm s$^{-1}$ on a range of timescales upward of about 200 days. 
The origin of these slow drifts is unclear. They could be a shift-like manifestation of solar activity. Secular drifts induced by the instrument are also a possibility. 

The Ljung-Box Q test \citep{1978Biom...65...297}
suggests that the shift-driven velocities remain weakly correlated at all autocorrelation lags up to at least 100 days \textcolor{black}{of observation.} 
Therefore,  it is likely that some activity-driven velocity components remain in $\mathbf{v}_\perp$, but \textcolor{black}{as the upper-right panel of Fig.~\ref{fig:velHel} shows,} they are substantially reduced relative to the original, observed velocity time series. \textcolor{black}{The shape-driven signals are, as expected, strongly correlated.} This offers improved detection prospects for small planetary-orbit signals at periods of tens to hundreds of days.

In Fig.~\ref{fig:periodogramHel} we show periodograms (in terms of the best-fit semi-amplitude of a sinusoid as a function of its period) for radial velocities measured with the data-reduction system, transformed to the heliocentric reference frame. 
The periodogram of the raw velocities shows numerous candidate signals with semi-amplitudes of order 0.4~m~s$^{-1}$, particularly between 13 and 26 days, close to the solar synodic rotation period and its first harmonic. The {\sc scalpels} projection shows a very similar pattern of semi-amplitudes. 

These peaks are strongly suppressed in the semi-amplitude periodogram of the shift-driven radial velocities (Fig.~\ref{fig:periodogramHel}, bottom trace), which shows no strong frequency structure. 
This is important, since the background level of the periodogram peaks in the cleaned timeseries with no planets present, effectively sets the sensitivity for detecting planets after applying {\sc scalpels} to clean the velocity measurements. 
Peaks with amplitudes greater than 0.30~m~s$^{-1}$ are seen in the shift-driven radial velocities at \textcolor{black}{$P=191.47$ and 30.45 days. Their amplitudes are reduced to 0.320 and 0.339 m~s$^{-1}$ respectively}, nearly a factor of two less than those found in the observed velocities measured with the data-reduction system. We note that the excess of power at around 200 days is commensurate with the average interval between cryostat warm-ups.

\section{Algorithm tests with planets injected into solar observations}
\label{sec:injtests}

We now turn to the problem of determining the impact of the {\sc scalpels} signal separation on detection thresholds when weak planetary signals are present. We begin by injecting four periodic shift signals into the heliocentric CCF timeseries, using equation~\ref{ccfShift} to shift the rows by the small amounts required. The periods of these signals were well-spaced in log period, at non-integer periods of 7.142, 27.123, 101.543 and 213.593 days. The 27.1-day period was chosen deliberately to be close to the solar synodic rotation period. The injected signals are sinusoidal in form, with semi-amplitudes $K=40$ cm s$^{-1}$. This is less than the amplitude of the strongest signals arising from solar variability in the upper trace of Fig.~\ref{fig:periodogramHel}, but greater than the activity-driven signals remaining after subtracting the {\sc scalpels} projection from the radial-velocity measurements. For a 1 M$_{\odot}$ star the corresponding planet masses are 1.2, 1.9, 2.9 and 3.7 M$_\oplus$.  

Before proceeding, we must consider the methodology used to extract velocities from the shifted CCFs and to estimate their precision from the covariances in the rows and columns of the CCF. The reader is referred to Appendix~\ref{sec:CCFexpansion} for the details of this methodology.

\begin{table*}
    \caption{Frequencies, periods and semi-amplitudes of the strongest signals in the periodograms of raw and shape-corrected apparent velocities. The first two columns give the periods and semi-amplitudes of the four injected signals. Columns 3, 4 and 5 give the periods, and semi-amplitudes and uncertainties of all peaks with amplitudes greater than 33~cm~s$^{-1}$ in the periodogram of residual velocities remaining after subtraction of the {\sc scalpels} projection, as in a blind RV search. 
    The final three columns give the same information, from simultaneous modelling of CCF shape changes and planetary motion, made with prior knowledge of the four injected periods, as described in  (Section~\ref{sec:fitsimul}).   
}
    \centering
    \begin{tabular}{cccccccc}
    \hline\\
\multicolumn{2}{c}{Injected} & \multicolumn{3}{c}{Velocities from residual CCF} & \multicolumn{3}{c}{Velocities from simultaneous fit} \\
$P$ & $K$ & $P$  & $K$  & $\sigma_K$  & $P$ & $K$ & $\sigma_K$ \\
(day) & (m s$^{-1}$) & (day) & (m s$^{-1}$) & (m s$^{-1}$) & (day) & (m s$^{-1}$) & (m s$^{-1}$) \\
\hline\\
   7.142 & 0.400 &   7.144 & 0.443 & 0.064 &   7.142 & 0.451 & 0.065 \\
  27.123 & 0.400 &  27.133 & 0.426 & 0.066 &  27.123 & 0.430 & 0.066 \\
 101.543 & 0.400 & 100.625 & 0.448 & 0.064 & 101.543 & 0.426 & 0.066 \\
 213.593 & 0.400 & 215.219 & 0.395 & 0.065 & 213.593 & 0.427 & 0.069 \\
 -- & -- & 185.101 & 0.435 & 0.065 & -- & -- & -- \\ 
 \hline\\
    \end{tabular}
    \label{tab:peaks}
\end{table*}

\begin{figure}
    \centering
    \includegraphics[width=\columnwidth]{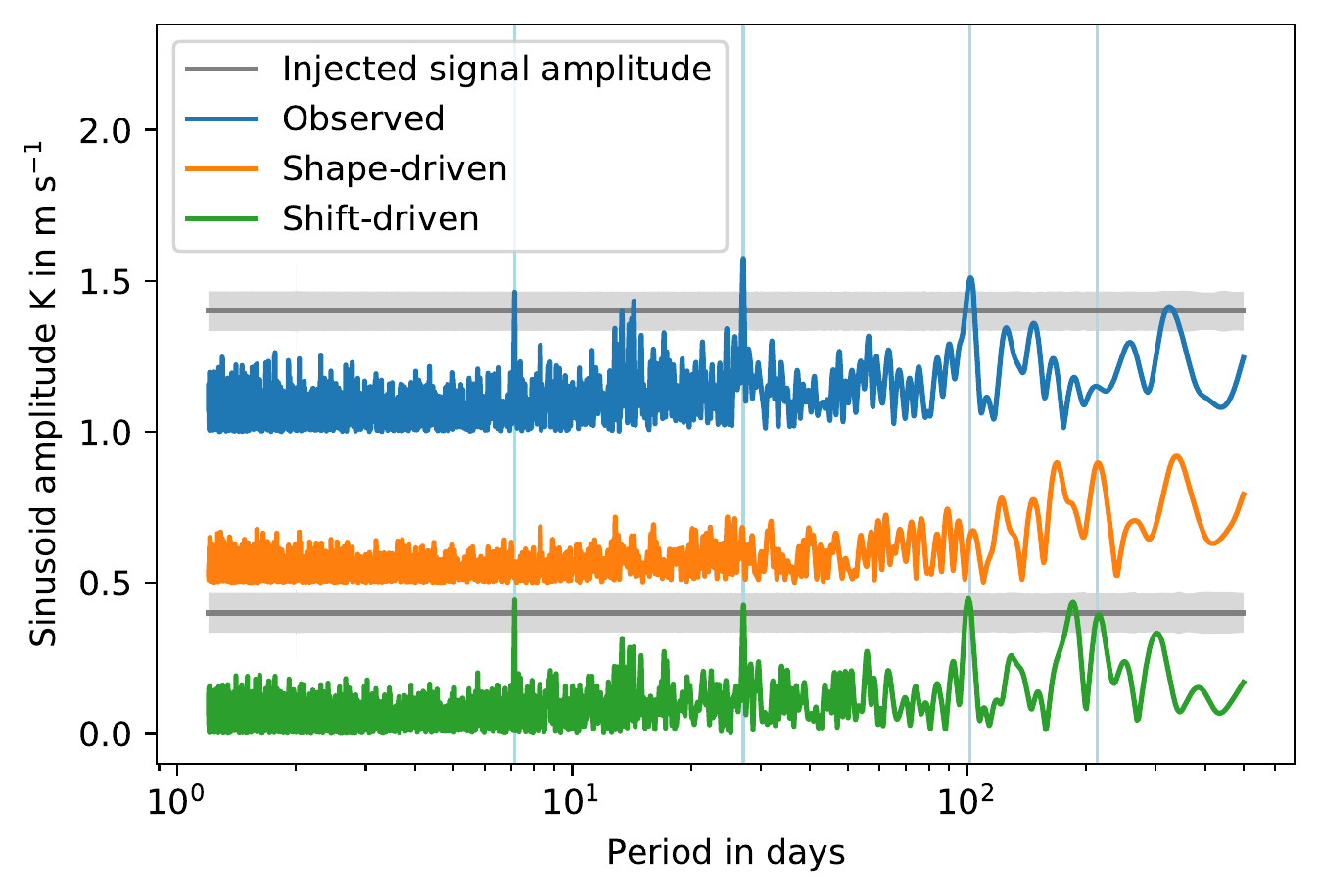}
    \caption{Periodograms of velocities derived from the heliocentric solar CCFs when four sinusoidal signals of 40 cm s$^{-1}$ have been injected at the periods denoted by the vertical blue lines. The traces are as defined in the caption of Fig.~\ref{fig:periodogramHel}. The uncertainty in the amplitude of the sinusoid is almost independent of period. Its 1-$\sigma$ limits are indicated by the shaded region around the horizontal grey lines showing the amplitude of the injected signal.
    The four dominant peaks in the lower trace indicate successful recovery of all four signals, at amplitudes that are consistent with the injected values, and whose scatter is consistent with the expected uncertainty.}
    \label{fig:recovery}
\end{figure}

\subsection{Recovery of weak injected planet signals}

Following signal injection, radial velocities were again measured from the CCF time series using equation~\ref{eq:getvel}.

Fig.~\ref{fig:recovery} shows the periodograms obtained from these velocities, from the {\sc scalpels} projection of the shape-driven velocity component, and from the differences between them representing pure shifts.

The periodogram of the velocities measured from the shifted CCFs does not enable us to distinguish clearly between the injected signals and RV variability intrinsic to the Sun or the instrument.
\textcolor{black}{The injected signals at 7.1 days, 27 days and 100 days are detected fairly unambiguously, but the 213-day signal is suppressed and there are also many false detections of amplitude comparable to the injected signals.} 

The periods, semi-amplitudes and uncertainties of the five strongest signals recovered from the periodogram of the shift-driven velocities after subtraction of the shape-driven model are listed in columns 3, 4 and 5 of Table~\ref{tab:peaks}. \textcolor{black}{Among these, four of the five strongest signals are very close to the frequencies of the injected planet signals. The mean of their semi-amplitudes is $0.428\pm 0.01$~m~s$^{-1}$, somewhat above the expected sample uncertainty of the injected values. They dominate over all residual variability and zero-point jitter signals except for a spurious 0.435~m~s$^{-1}$ signal at $P=185.1$ d. This latter period is so close to half a year that it is likely to arise from an as-yet unidentified effect of observing the Sun from Earth, which would not be expected to affect exoplanet searches.}

\subsection{Simultaneous modelling of stellar variability and planetary motion}
\label{sec:fitsimul}

The amplitudes of the injected planet signals do not appear to be strongly attenuated in the upper trace of Fig.~\ref{fig:recovery}, but they are buried in a forest of activity-related peaks of similar amplitude. In the bottom trace they stand out above the suppressed activity signals. Their amplitudes could nonetheless be affected by activity if the planet signals themselves contaminate the {\sc scalpels} projection. This could occur if the injected shift signals are not perfectly orthogonal to all elements of $\mathbf{U}_A$, and hence partly absorbed in the {\sc scalpels} projection. The dataset has a finite length, so irregularly-sampled superpositions of keplerian signals will not be perfectly orthogonal to any randomly-chosen vector in the same space. Moreover, a periodogram fits only a single sinusoid per frequency sample, so that cross-talk between multiple signals can lead to incorrect amplitude estimates.

The orbital perturbations of any planet and the {\sc scalpels} projection process must therefore be modelled self-consistently for the signal separation to recover their semi-amplitudes as reliably as possible. Once the periods of candidate signals have been determined -- either through prior knowledge of transits, or via periodogram search(es) -- parameter estimation and signal separation can be achieved in a single linear calculation. 

For a set of $n$ planet signals, the net orbital velocity vector \textcolor{black}{$\mathbf{v}_{\rm orb}$ can be modelled as} the product of a set of coefficient pairs $\mathbf{\theta}_{\rm orb} = \{A_1, B_1, \cdots, A_n, B_n\}$ with an array of time-domain function pairs
$\mathbf{F} = \{\cos\omega_1 t_j, \sin\omega_1 t_j, \cdots, \cos\omega_n t_j, \sin\omega_n t_j\}$, $\omega_k$ being the orbital frequency of the $k$th planet:
\textcolor{black}{
\begin{equation}
\mathbf{v}_{\rm orb} = \mathbf{F}\cdot \mathbf{\theta}_{\rm orb},
\end{equation}
For simplicity we assume circular orbits here, though eccentric orbits could in principle be fitted with periodic signals including additional Fourier components.}

\textcolor{black}{The complete model of the radial-velocity data is then the sum of the model orbital velocity variations and the shape-driven velocity variations. The only unknowns are the amplitudes and phases of the orbital basis functions $\mathbf{F}$. We can solve for these using the method of least squares.}

\textcolor{black}{The simultaneous solution involves computing the shape-driven variations from the difference between the observed radial velocities and the model velocities. 
In the projection-operator language of Section~\ref{sec:scalpels}, we solve for the vector $\mathbf{\theta}_{\rm orb}$ that minimises 
$\chi^2=(\mathbf{P}_\perp\cdot\delta\mathbf{v}^T)
    \cdot
    \Sigma^{-1}
    \cdot
    (\mathbf{P}_\perp\cdot\delta\mathbf{v})$, where $\delta\mathbf{v}\equiv\mathbf{v}_{\rm obs}-\left<v_{\rm obs}\right>-\mathbf{F}\cdot \mathbf{\theta}_{\rm orb}$.}
    
\textcolor{black}{Defining $\mathbf{v}_\perp=\mathbf{P}_\perp\cdot\mathbf{v}_{\rm obs}$ and $\mathbf{F}_\perp=\mathbf{P}_\perp\cdot\mathbf{F}$, the goodness of fit is quantified by
\begin{equation}
    \chi^2=(\mathbf{v}_\perp-\mathbf{F}_\perp\cdot\mathbf{\theta}_{\rm orb})^T
    \cdot
    \Sigma^{-1}
    \cdot
    (\mathbf{v}_\perp-\mathbf{F}_\perp\cdot\mathbf{\theta}_{\rm orb}),
\end{equation}
which is minimised by solving for $\mathbf{\theta}_{\rm orb}$:
\begin{equation}
(\mathbf{F}_\perp^T \cdot\mathbf{\Sigma}^{-1}\cdot\mathbf{F}_\perp)\cdot\mathbf{\theta}_{\rm orb} =
\mathbf{F}_\perp^T \cdot\mathbf{\Sigma}^{-1}\cdot\mathbf{\mathbf{v}_\perp}.
\label{eq:solvetheta}
\end{equation}
}

\textcolor{black}{The log likelihood of the data given the model is 
\begin{eqnarray}
    \ln\mathcal{L} &= 
    &-\frac{1}{2}(\mathbf{v}_\perp-\mathbf{F}_\perp\cdot\mathbf{\theta}_{\rm orb})^T\cdot\mathbf{\Sigma}^{-1}\cdot(\mathbf{v}_\perp-\mathbf{F}_\perp\cdot\mathbf{\theta}_{\rm orb})\nonumber\\
    & &-\frac{1}{2}\ln|\mathbf{\Sigma}|-\frac{m}{2}\ln(2\pi)\nonumber\\
    &= 
    &-\frac{1}{2}(\mathbf{v}_\perp-\mathbf{F}_\perp\cdot\mathbf{\theta}_{\rm orb})^T\cdot\mathbf{\Sigma}^{-1}\cdot(\mathbf{v}_\perp-\mathbf{F}_\perp\cdot\mathbf{\theta}_{\rm orb})\nonumber\\
    & & -\frac{1}{2}\left(\sum_{j=1}^{m}\ln\mathbf{\Sigma}_{jj}\right)-\frac{m}{2}\ln(2\pi)
    \label{eq:likelihood}    
\end{eqnarray}}

\begin{figure}
    \centering
    \includegraphics[width=\columnwidth]{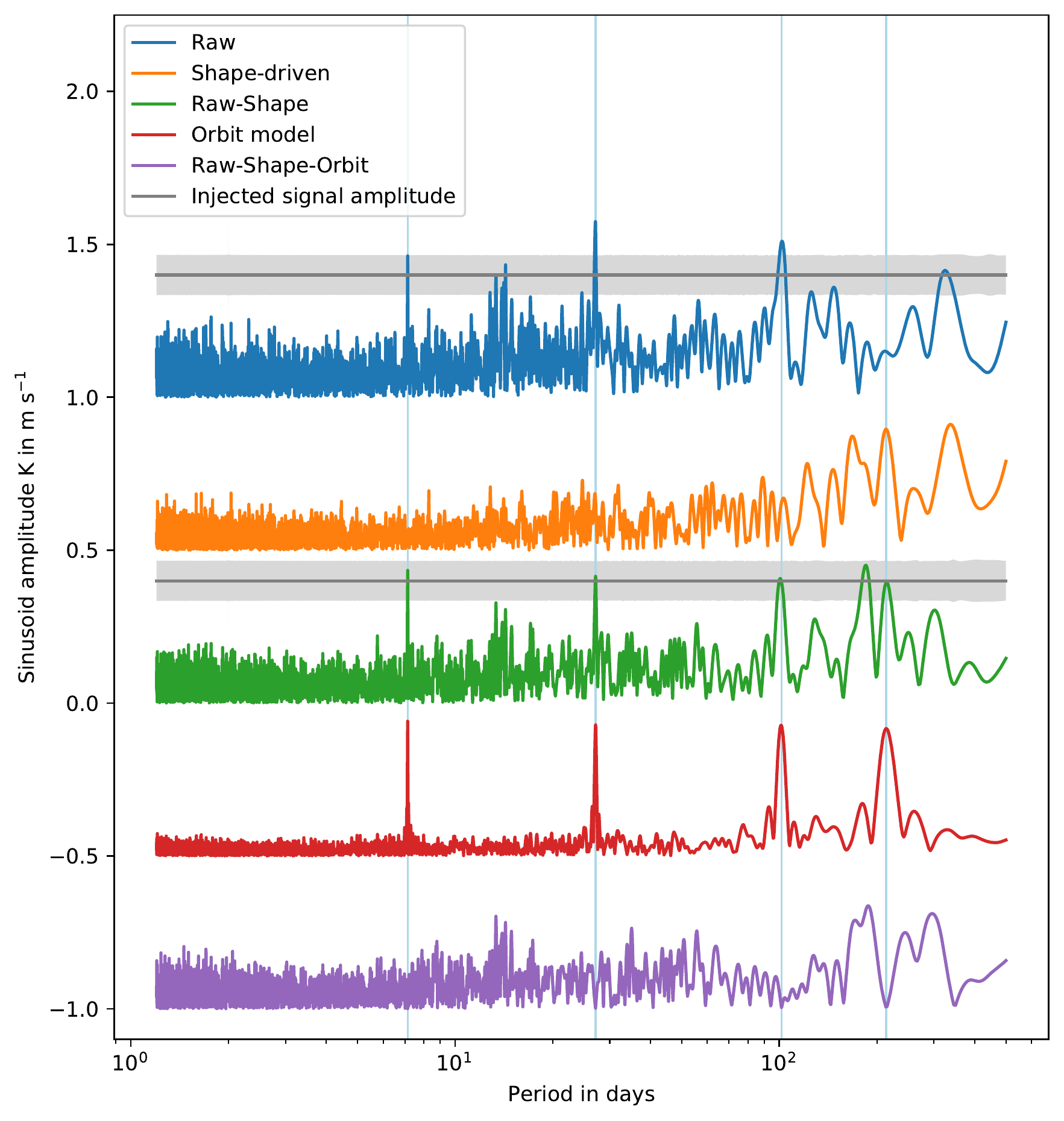}
    \caption{As for Fig.~\ref{fig:recovery}, but for the case where signal separation is performed simultaneously with orbit fitting given prior knowledge of the orbital periods. The middle (green) periodogram shows the difference between the observed and shape-driven velocities. The fourth (red) trace shows the periodogram of the fitted model of the four orbital signals. The bottom (magenta) trace shows the residuals after subtraction of both the shape-driven and orbital RV models. The scatter in the amplitudes of the four dominant peaks in the middle trace is consistent with the expected uncertainty, and their mean amplitude is unbiased relative to the injected values.}
    \label{fig:recovery_simult}
\end{figure}

\begin{figure}
    \centering
    \includegraphics[width=\columnwidth]{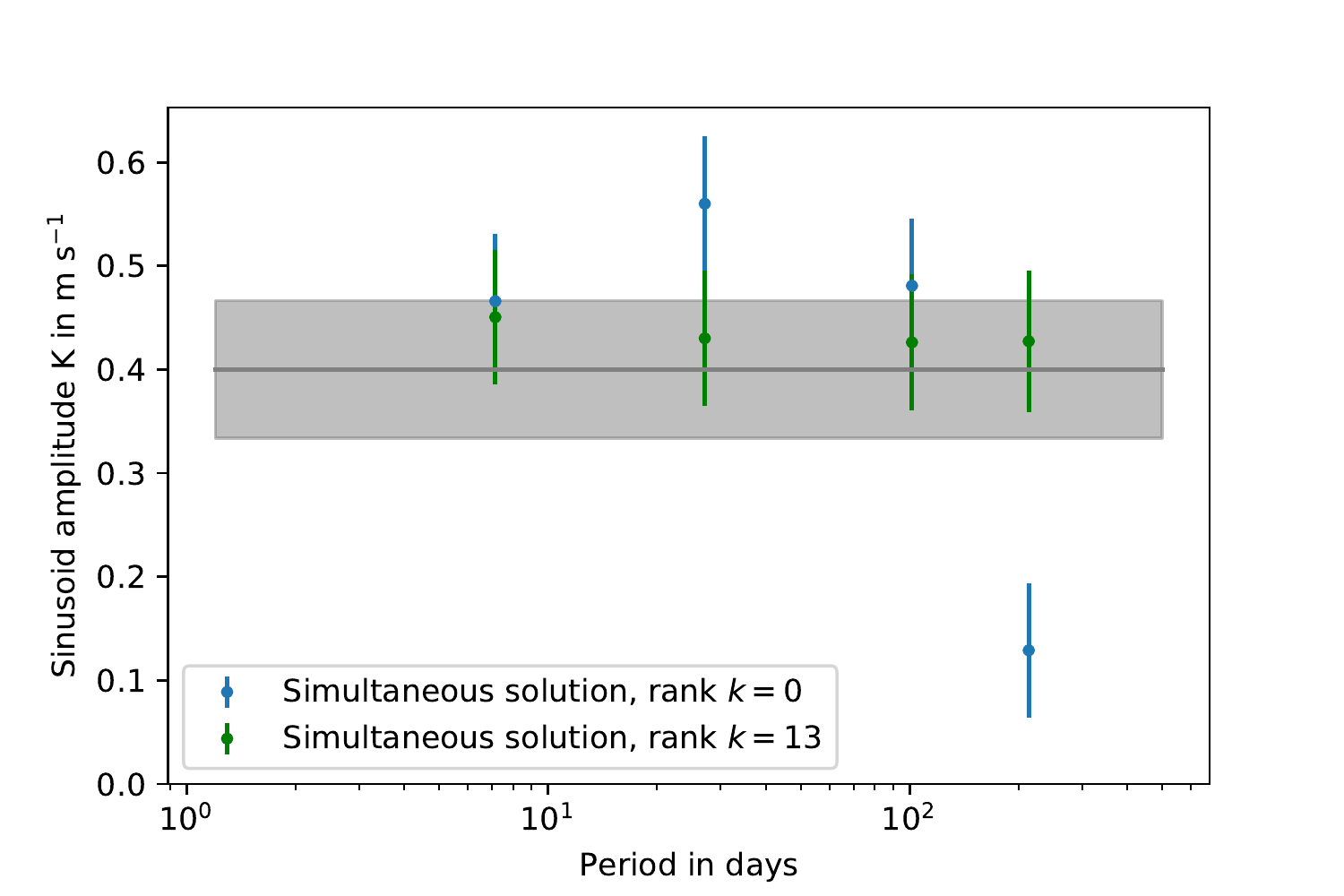}
    \caption{\textcolor{black}{The recovered amplitudes of the injected signals are shown in green for the simultaneous {\sc scalpels} fit with $k_{\rm max}=13$, and in blue for simultaneous sinusoidal fits to the uncorrected original RV data ($k_{\rm max}=0$) at the injected periods. The grey line and band show the original signal level and the formal 1-$\sigma$ uncertainty on the recovered amplitude. The result demonstrates clearly the improvement in consistency and fidelity in the recovered signal amplitudes when {\sc scalpels} is used.}}
    \label{fig:amp_simultfit}
\end{figure}

\textcolor{black}{Here we use the simplifying assumption that the RV measurements are uncorrelated. The covariance matrix is diagonal and its log determinant is $\sum_{j=1}^{l}\ln\mathbf{\Sigma}_{jj}$. The diagonal elements $\Sigma_{jj}={\rm Var}(\mathbf{v}(t_j))$ are calculated using equations \ref{eq:getvar} and \ref{eq:covmatrix}. If the radial-velocity data are sufficiently densely sampled, a time-dependent covariance model with a kernel incorporating the stellar rotation period and active-region lifetime could also be included -- see, e.g. \cite{2020ApJ...905..155G}. 
For convenience, we summarise the algorithm in Appendix~\ref{sec:appsimul}. }

In Table \ref{tab:peaks}, columns 6, 7 and 8 list the amplitudes and uncertainties of the sinusoidal signals recovered from the data at the known periods of the injected signals. \textcolor{black}{The standard deviation of the four individual recovered semi-amplitudes is $\sigma = 0.010$~m~s$^{-1}$. The sample mean and standard deviation ($\sigma/\sqrt{4}$) of the signal amplitudes are $0.433\pm 0.005$~m~s$^{-1}$, again somewhat above the injected value. The four individual signals deviate from the injected values by amounts that are consistent with their individual estimated 0.066~m~s$^{-1}$ semi-amplitude uncertainties.} 

The improvement in signal separation is also apparent from Fig.~\ref{fig:recovery_simult}. \textcolor{black}{The top two traces are almost the same as those in Fig.~\ref{fig:recovery}, but the balance of the signal separation is changed by the explicit modelling of the orbital motion at the known periods.} The periodogram of the fitted orbital model illustrates the apparent signal attenuation that can occur when fitting multiple signals with a single sinusoid. The final residuals are very similar to those of Fig.~\ref{fig:periodogramHel}.

\textcolor{black}{The precision of the recovered semi-amplitudes is poorer when the velocities are left uncorrected for profile-shape variations, by setting the dimension $k_{\rm max}$ of the null space to zero. If sinusoids are fitted to the raw $\mathbf{v}_{\rm obs}$ at the same four periods, we obtain semi-amplitudes 0.466, 0.560, 0.481 and 0.129 ms $^{-1}$, whose sample mean and standard deviation are $ 0.409\pm 0.083$~m~s$^{-1}$. Fig.~\ref{fig:amp_simultfit} shows clearly the improvement in fidelity of the recovered amplitudes when the optimal shape model is applied.}

\textcolor{black}{The formally-propagated 0.066~m~s$^{-1}$ uncertainties in the semi-amplitudes are  $\sim 1.6\sigma_{\rm v} /\sqrt{N_{\rm obs}}$ for $N_{\rm obs}=853$ if we adopt a single-measurement precision of $\sigma_{\rm v}\simeq 1.25$~m~s$^{-1}$ based on the RMS scatter in the heliocentric solar velocities after removal of the shape perturbations (see Section~\ref{sec:scalpels_solar}).The RV amplitude precision appears to scale as expected for uncorrelated random variables to a level $\sim 20$ times better than the single-measurement precision.} 

The average of the recovered semi-amplitudes shows no evidence of bias relative to the injected value when signal separation is performed with prior knowledge of the orbital period, as is the case with transiting planets. 
Thus, the RV semi-amplitudes inferred from the cleaned velocities can be significantly more reliable than RV semi-amplitudes inferred from original velocity measurements. 

We conclude that the {\sc scalpels} method succeeds in reducing correlations between apparent velocities due to stellar variability, based solely on line shape changes and without making use of time-domain information. 
This decoupling from time-domain information allows planet signals to be recovered with good fidelity even when they fall close to the stellar rotation period, as is the case with the 27-day signal injected here.

\section{Discussion}
\label{sec:discuss}

\subsection{Summary}
\label{sec:summary}
We have presented a new algorithm for extracting precise radial velocity estimates from high-resolution spectroscopic planet surveys.  
The algorithm begins with a list of CCFs for each observation epoch, constructs a reduced-rank representation of stellar variability and reconstructs CCFs which have been cleaned of most stellar variability.  
We demonstrated the algorithm using observations of the solar spectrum from HARPS-N.  
We verified and validated that the algorithm can accurately detect multiple simulated planets injected into solar observations, spanning a wide range of orbital periods.  

\subsection{Planets injected into solar observations}
\label{sec:summarysolar}
Our algorithm recovered the radial velocity signals of injected planets with semi-amplitudes of just 40 cm~s$^{-1}$ with a signal-to-noise ratio of \textcolor{black}{6} based on a timeseries of \textcolor{black}{853} CCF measurements.  
\\
The semi-amplitude uncertainty of \textcolor{black}{$6.6$ cm s$^{-1}$} implies that it is possible for intensive Doppler spectroscopy campaigns to measure the masses of transiting planets (i.e., with well-measured orbital period and inclination) at {\bf $\sim 15\%$} precision for RV semi-amplitudes as small as $K=40$ cm s$^{-1}$ for a solar twin, even if there are multiple planets spanning a wide range of orbital periods.   
\\
The  precision with which we can measure velocity semi-amplitudes is insensitive to the velocity semi-amplitude of the planet (for small velocity semi-amplitudes).  
Thus, our results suggest that intensive Doppler spectroscopy campaigns could detect and measure the masses (times sine of inclination angle) to $6\sigma$ precision for planets with RV semi-amplitudes of \textcolor{black}{$\sim 60$ cm s$^{-1}$}, if there were independent evidence for a planet at a given period (e.g., transit, direct imaging) orbiting \textcolor{black}{a sufficiently bright ($V\simeq 6$)} Sun-like star.
\\

\subsection{Areas for Further Research}
\label{sec:future}
The algorithm presented successfully mitigates the impact of solar variability.  Nevertheless, there are multiple lines for additional research that are likely to lead to further improvements in the Doppler sensitivity.

\subsubsection{Simulations using realistic observing cadence}
For planet injection tests in \S\ref{sec:injtests}, we used CCFs from HARPS-N solar observations taken on 853 of a total 886 
days.  
The observations span 1681 days, and over 70\% of the observations are spaced by just one day.  While there are seasonal shutdowns each year and gaps of up to two weeks, this observing cadence is more favourable than that of most targets of Doppler planet surveys.
We recommend that future simulations of Doppler planet surveys explore how the planet detection sensitivity depends on observing cadence in the presence of intrinsic stellar variability and after incorporating advanced methods for mitigating stellar variability such as presented here. \citet{2018MNRAS.479.2968H} have carried out such a study for the HARPS-3 project, which will produce stellar datasets of comparable duration and quality to the solar dataset examined here, albeit with seasonal gaps.
A similar sensitivity study using HARPS-N solar data has been published recently by \citet{2020arXiv200805970L}, who concluded that a decade or more of observations could be needed to achieve a 5$\sigma$ detection of a 50 cm s$^{-1}$ signal with a period of 225 days, given an instrumental white-noise uncertainty of 80 cm s$^{-1}$ and a perfect model of activity-driven RV variability. \citet{2020arXiv200513386H} analysed an 8-year sequence of synthetic solar RVs derived from SDO images, with a six-month duty cycle and decorrelation against hemispherically-averaged magnetic field. They succeeded in recovering an injected signal with $K=30$ cm s$^{-1}$ and $P=300$ days. Our results, using real data but an idealised observing pattern, give similar cause for optimism. 

\subsubsection{Granulation}
The CCF timeseries used for verifying our algorithm was based on \textcolor{black}{15-minute} daily averages of the HARPS-N solar CCFs.  
Each exposure time is 300 seconds, yielding a signal-to-noise ratio in the range $250 < {\rm SNR} < 400$. This is comparable to that of exoplanet observations for a host star of magnitude 5.5 using the same exposure time.  
\textcolor{black}{Combining three contiguous such exposures in a 15-minute block} is reasonably effective at averaging-out the side effects of $p$-mode pulsations which occur on a $\sim 5$ minute timescale \citep{2019AJ....157..163C}.  
Modern exoplanet surveys typically choose an exposure time of at least 15 minutes, averaging sub-exposures if necessary to average out spectral variability due to pulsations and avoid saturation. Such exposures are repeated at intervals of 2-3 hours to mitigate the effects of granulation. \textcolor{black}{Our use of a single 15-minute block per day is likely to be less effective at eliminating intrinsic stellar variability due to the granulation pattern, as studied in detail by \citet{2015A&A...583A.118M}.  We tested this by averaging all data satisfying the data-selection constraints in Section 2 throughout each day of observations, rather than down-selecting to a single 15-minute block. The longer daily baseline reduced the day-to-day scatter in $\mathbf{V}_\perp$ from 1.25~m~s$^{-1}$ to 1.08~m~s$^{-1}$. This is less than the improvement expected if photon counts were the limiting noise source, but consistent with the improvement expected through averaging over velocity fluctuations arising from photospheric granulation noise.}

\textcolor{black}{In principle, granulation might imprint on the stellar spectrum differently than variability on the timescales of magnetic activity \citep{2019Geosc...9..114C}. It is likely that some of the 13 modes of radial-velocity variability found in our data may be attributable to granulation.  }

\subsubsection{Searching for low-mass planets with outer giants}

Some complexities of planetary-system architecture are beyond the scope of this initial study, but merit further investigation. For example, signals from terrestrial planets with sub-m~s$^{-1}$ amplitudes may be superposed on much larger signals of giant planetary companions at longer orbital periods. If the giant's orbit is well-characterised, it can be included directly in the model. Indeed the barycentric data considered here contain such a signal, with the synodic period of Jupiter. Rather than simply work in the heliocentric frame, we repeated the analysis of Section~\ref{sec:fitsimul} by injecting the same four signals into the barycentric CCFs.  To mimic the situation where the giant's orbit is incomplete or poorly-determined, we performed a GP regression with a squared-exponential covariance kernel to smooth out variations on timescales longer than 300 days. All four of the injected signals were recovered successfully at the correct amplitudes. 

More complex cases might involve transiting systems in which a temperate Earth-sized object is accompanied by a few strongly interacting compact short-period companions. In this case, gravitational interactions might make the fully linear algebra approach of the method less effective. As shown in Fig. 9, however, signal separation and detection is reasonably effective even in a ``blind-search" scenario where simultaneous linear fitting is not possible. Furthermore, even a nonlinear parametric model of the orbital reflex motion can be included as a fitting function and its parameters optimised together with the coefficients of ACF basis vectors in an MCMC scheme, at the cost of some additional computational overheads.

\subsubsection{Reconstructing full spectra}
In this paper, we reconstructed the shape-driven velocity signal using a reduced-rank representation and scores derived from the ACF of the CCFs.  
In principle, the same approach could be used to reconstruct the CCFs themselves or even the full spectra.  
Then, the reconstructed spectra could become objects for further analysis, e.g., performing line-by-line analysis \citep{2018A&A...620A..47D}.  
We anticipate that a reconstruction of the spectra would likely lead to more significant deviations from the observed spectra than those we found when reconstructing the velocities alone from the ACF. This is because the CCF, from which the ACF is derived, combines information from a large list of spectral lines, even though different lines are known to respond differently to stellar variability \citep{1981A&A....96..345D,1988ApJ...334.1008T}.
Furthermore, variability of the continuum contributes very little to the CCF.
Future studies could analyze the residual spectra to gain insights into more subtle ways in which stellar variability manifests in the observed spectra \citep[e.g.,][]{2017MNRAS.468L..16T,Davis2017,2018A&A...620A..47D,2018AJ....156..180W,2020A&A...633A..76C}..

\subsubsection{Multiple CCF masks}
The results of full-spectrum reconstruction could inform the development of additional stellar variability indicators and/or the design of CCF masks.
As noted above, summarizing the spectrum with a single CCF averages over the different responses of lines which are more or less sensitive to stellar activity.  
For this study, we have used the CCF generated by the HARPS-N DRS. 
However, one could design multiple CCF masks to compute multiple CCFs at each observation epoch.
The choice of mask could be based on astrophysical insights or machine learning approaches.  
Either way, each mask would include spectral lines exhibiting common patterns of response to stellar variability, so as to provide greater sensitivity for recognizing line shape variations. 

Once CCFs have been computed using multiple masks, the ACF can be computed for each CCF-mask pair. The underlying algorithm presented in Section~\ref{sec:scalpels} can be applied to compute time-domain subspaces either separately for each mask or simultaneously.
The resulting velocities derived from each mask could be analyzed separately for diagnostic purposes and then combined for inference \citep[e.g.,][]{Zechmeister2017}.  
The use of multiple masks naturally leads to multivariate timeseries for estimates of the velocity and stellar variability indices.  

\subsubsection{Time-domain modelling}
Stellar variability generally distinguishes itself from planet-induced orbital motion both in the wavelength domain (e.g., relative depths of lines, line shapes) and in the time domain (e.g., deviations from a strictly Keplerian signal). We caution, however, that subtler effects such as activity-driven changes in the stellar gravitational redshift may manifest as pure shifts at amplitudes up to 10 cm s$^{-1}$ \citep{2012MNRAS.421L..54C}.
In this paper, we have described an improved algorithm for measuring radial velocities in the presence of stellar variability that does not make use of time-domain information. That is, the computation of $\mathbf{v}_\parallel$ depends only on the ensemble of measured CCFs, but not on the times at which the observations were taken. For the purpose of demonstrating our algorithm, we have used traditional maximum likelihood estimation based solely on the cleaned velocities.  
Our algorithm naturally produces additional stellar variability indicators (i.e., $\mathbf{U}_A(t)$ and $\mathbf{U}_C(t)$)) that could be modelled simultaneously with the velocities, following methods developed by \citet{2015MNRAS.452.2269R} and generalized in  \citet{Jones2017}.

We recommend that future simulations explore the potential for further improvements in the sensitivity to small planets by combining algorithms for utilizing information in the wavelength and time domains.  

\subsubsection{Integrating into Doppler planet survey toolboxes}
The algorithm developed in his paper can be implemented efficiently using a standard linear algebra toolbox.  
Thus, it can be readily integrated into existing or future software packages to analyze Doppler planet search observations.  
For example, one could inspect the posterior marginalized over all parameters except orbital period (i.e., a periodogram), similar to \citet{2015A&A...573A.101M}, but replacing the standard likelihood with equation~\ref{eq:likelihood}.
When considering observations of a star potentially hosting multiple planets, a periodogram of the reconstructed CCFs could be applied iteratively, removing one signal at a time.
Alternatively, one could apply sparse regression techniques, i.e.,  fit for the semi-amplitudes of many periodic signals simultaneously, while applying regularization to the semi-amplitude, so as to find a family of maximum likelihood solutions for each plausible number of planets. 

This could be implemented efficiently for an {\em a priori} unknown number of signals using either the alternating direction method of multipliers (ADMM) algorithm or the spectral projected-gradient algorithm \citep[as in ][]{2017MNRAS.464.1220H}.  
Once the putative orbital periods have been identified, then MCMC-based techniques can be used to perform parameter estimation \citep[e.g.,][]{Ford2006,Nelson2014} replacing the standard likelihood with our Eqn.\ref{eq:likelihood}.
Then, one could compare the Bayes factor or ratio of marginalized likelihoods (i.e., ``evidences'') for models with various numbers of planets.  
While there is still active research in finding efficient and robust algorithms for estimating Bayes factors for models with several planets, replacing the likelihood with Eq.~\ref{eq:likelihood} would be algorithmically straightforward and is expected to add only a modest additional cost.

\subsubsection{Design of Doppler planet surveys}
The observing strategy of Doppler planet surveys (e.g., number of observations per star, distribution of duration between observations) can have a significant impact on the sensitivity for detecting planets and its dependence on orbital period.  
Simulations of Doppler planet surveys have been used to inform survey design choices  \citep[e.g.,][]{Ford2008,2018AJ....156...82C,2018AJ....156..255B,2018MNRAS.479.2968H}.

Previous studies have typically ignored or adopted simplistic models for intrinsic stellar variability (e.g., assuming that a fixed fraction of spurious velocities due to stellar variability can be corrected).  
Our algorithm for simultaneously inferring the effects of stellar variability and planetary signals could be readily incorporated into more advanced survey simulations incorporating explicit models of the effects of stellar activity such as that published recently by \citet{2019MNRAS.489.2555D}.

Now that there is a concrete and computationally efficient strategy for mitigating stellar variability, we recommend that future studies conduct new survey simulations to compare candidate Doppler planet survey strategies.  
We anticipate that our method will significantly increase the value of observing the same star many times, since we substantially reduce the correlation of inferred velocities at different times.
On the other hand, it may be that the improved ability to mitigate stellar activity reduces the importance of obtaining a dense set of observations over the rotation period or active region lifetime.  
Monte Carlo simulations are necessary to understand the interaction of these two effects and the implications for future planet surveys.  
In addition to informing survey strategies that do not make use of information from previous observations, our algorithms could be folded into adaptive scheduling algorithms that maximize a merit function (or minimize a cost function), so as to maximize the efficiency of a Doppler survey \citep[e.g.,][]{Ford2008}.

\section*{Acknowledgements}
This article is a result of collaborative scholarly efforts from the residency of the Research Group on Big Data and Planets at the Israel Institute for Advanced Studies.
We acknowledge valuable discussions with Tsevi Mazeh, Eric Feigelson, Shay Zucker, Lev Tal-Or and Dovi Poznanski.  
\textcolor{black}{We thank the anonymous referee for numerous insightful suggestions that led to major improvements, including the use of leave-one-out cross-validation and streamlining of the mathematical approach used in Section~\ref{sec:fitsimul}.}
ACC acknowledges support from
the Science and Technology Facilities Council (STFC) consolidated
grant number ST/R000824/1 and UKSA grant ST/R003203/1.
EBF acknowledges support from NSF award \#1616086, NASA grant \#80NSSC18K0443, and Heising-Simons Foundation Grant \#2018-0851.
This work was supported by a grant from the Simons Foundation/SFARI (675601, E.B.F.).
E.B.F. acknowledges the support of the Ambrose Monell Foundation and the Institute for Advanced Study.
EBF acknowledges support from the Penn State Eberly College of Science, Department of Astronomy \& Astrophysics, Institute for Computational \& Data Sciences, the Center for Exoplanets and Habitable Worlds, the Center for Astrostatistics. 
DFP acknowledges NASA award number NNX16AD42G.
This work was performed partly under contract with the California Institute of Technology (Caltech)/Jet Propulsion Laboratory (JPL) funded by NASA through the Sagan Fellowship Program executed by the NASA Exoplanet Science Institute (RDH).
AM acknowledges support from the senior Kavli Institute Fellowships.
XD is grateful to The Branco Weiss Fellowship--Society in Science for its financial support. 
HMC acknowledges financial support from the National Centre for Competence in Research (NCCR) PlanetS, supported by the Swiss National Science Foundation (SNSF), as well as a UK Research and Innovation Future Leaders Fellowship. This project has received funding from the European Research Council (ERC) under the European Union's Horizon 2020 research and innovation programme (grant agreement No 851555)
This work has been carried out in the framework of the National Centre for Competence in Research \emph{PlanetS} supported by the Swiss National Science Foundation (SNSF).
The authors gratefully acknowledge the support of the International Team 453 by the International Space Science Institute (Bern, Switzerland).
Based on observations made with the Italian {\it Telescopio Nazionale Galileo} (TNG) operated by the {\it Fundaci\'on Galileo Galilei} (FGG) of the {\it Istituto Nazionale di Astrofisica} (INAF) at the  {\it Observatorio del Roque de los Muchachos} (La Palma, Canary Islands, Spain). 
The HARPS-N project has been funded by the Prodex Program of the Swiss Space Office (SSO), the Harvard University Origins of Life Initiative (HUOLI), the Scottish Universities Physics Alliance (SUPA), the University of Geneva, the Smithsonian Astrophysical Observatory (SAO), and the Italian National Astrophysical Institute (INAF), the University of St Andrews, Queen’s University Belfast, and the University of Edinburgh.
%
%
The citations in this paper have made use of NASA's Astrophysics Data System Bibliographic Services.  
%


\section*{Data availability}
The HARPS-N solar data products for the first 3 years of the period described in this paper are publicly available through
the DACE platform (\url{https://dace.unige.ch}) developed in the frame of PlanetS. All other research data underpinning this publication (\url{https://doi.org/10.17630/9ec51274-cbea-445f-b188-112f4734c6e9}) and the {\sc python} code and notebook used to prepare all diagrams in this paper (\url{https://doi.org/10.17630/957d6320-1969-43c8-b6bb-a8ae53d1f657}) will be made available through the University of St Andrews Research Portal.





\appendix

\section{Calculation of profile derivatives}
\label{sec:appA}

Derivatives of the rows of the CCF with respect to velocity are required for applying radial-velocity shifts to the rows of the CCF in both correction from the barycentric to the heliocentric reference frame (Section~\ref{sec:barytohel}), and for injecting synthetic orbital reflex motion signals (Section~\ref{sec:injtests}). 

The numerical derivatives are calculated with a simple differencing scheme:
\begin{equation}
    \mathbf{C'}(v_i,t_j)=\frac{\mathbf{C}(v_{i+1},t_j)-\mathbf{C}(v_{i-1},t_j)}{2h}
    \label{eq:appdfdv}
\end{equation}
and
\begin{equation}
    \mathbf{C''}(v_i,t_j)=\frac{\mathbf{C}(v_{i+1},t_j) -2\mathbf{C}(v_i,t_j)+\mathbf{C}(v_{i-1},t_j)}{h^2}.
    \label{eq:appd2fdv2}
\end{equation}
Here $h$ is the velocity increment per CCF sampling interval in velocity.
The current HARPS-N DRS outputs CCFs onto a standard grid of $l=161$ velocities with $h=0.25$~km~s$^{-1}$.
Here $\mathbf{C}(v_i,t_j)$, $\mathbf{C'}(v_i,t_j)$ and $\mathbf{C''}(v_i,t_j)$ are estimates of the CCF and its derivatives, which we describe below.  

Simple numerical derivatives computed directly from the data have the undesirable property that they amplify noise.  
If the profile shape were time-invariant, this problem could be overcome by fitting the Taylor-series model derived from the mean profile, to a sequence of CCFs shifted by orbital reflex motion. 
However, magnetically-active regions rotating across the star's visible hemisphere distort the CCF profile \citep{2010A&A...519A..66M}. 
This time-variation in the shape of the profile alters the radial velocity measured by fitting a fixed profile or parametric function to the CCF. 
Currently, it is unclear whether the cross-terms between such intrinsic stellar variability and small Doppler shifts caused by planets are sufficiently small to be neglected.
Therefore, this paper attempts to estimate the derivative of the CCF at each observation time and focuses on solar data for which very high signal-to-noise data is available.  

Rather than estimating the derivative from a constant mean spectrum, we estimate the derivative of the CCF at each epoch $t_j$ based on a reduced-rank reconstruction of the CCF using the mean profile and the $k_{\rm max}=10$ leading principal components obtained by singular-value decomposition (SVD) of the full ensemble of CCFs. The optimal choice of $k_{\rm max}$ that best separates the effects of profile shape changes from the effects of oversampled photon noise is calculated as described in Appendix~\ref{sec:covmat}.

\section{Taylor-series velocity measurement from CCF}
\label{sec:CCFexpansion}

The radial velocities derived by the HARPS-N DRS are obtained by fitting a Gaussian profile to the activity-distorted CCF. Pure Doppler shifts simply displace the distorted profile in velocity.
Following the formulation of \cite{2001A&A...374..733B} for line shifts that are small compared to the line width, we measure Doppler shifts using a first-order Taylor-series approximation to the instantaneous CCF profile shape at the time $t_j$ of the $j$th observation. 
For shifts much smaller than the 0.25 km s$^{-1}$ velocity increment per CCF array element, the first derivative $\mathbf{C'}(v_i,t_j)$ of the instantaneous profile $\mathbf{C}$  suffices to measure the displacement of the mean-subtracted CCF $\mathbf{R}(v_i,t_j) = \mathbf{\mathbf{C}}(v_i,t_j) - \meanccf$ where $\meanccf$ is the time-averaged profile:
\begin{equation}
\mathbf{v}(t_j) = \frac{\mathbf{R}(v,t_j)^T \cdot \mathbf{\Sigma}^{-1}(t_j) \cdot \mathbf{C'}(v,t_j)}{\mathbf{C'}(v,t_j)^T \cdot \mathbf{\Sigma}^{-1}(t_j) \cdot \mathbf{C'}(v,t_j)}.
\label{eq:getvel}
\end{equation}
The derivatives $\mathbf{C'}(v_i,t_j)$ are calculated from a reduced-rank representation of $\mathbf{C}(v_i,t_j)$ as described in Appendix \ref{sec:appA}.
The covariance matrix $\Sigma$ is calculated as described in Appendix~\ref{sec:covmat} below.
The corresponding variances of the estimated radial velocities are given by
\begin{equation}
{\rm Var}(\mathbf{v}(t_j)) = \frac{1}
{\mathbf{C'}(v,t_j)^T \cdot \mathbf{\Sigma}^{-1}(t_j) \cdot \mathbf{C'}(v,t_j)}.
\label{eq:getvar}
\end{equation}

In order to ensure that the sampling of the CCF did not bias the scale of the derived velocities, we performed linear regression of both the DRS velocities and the Taylor-series velocities against the solar barycentric radial velocity computed with JPL {\sc horizons} \citep{1996DPS....28.2504G}. The scale factors differed from unity by 0.8 and 2.4 percent respectively. The barycentric radial velocities were scaled by these factors for all subsequent calculations, to ensure that the velocities were transformed correctly to the heliocentric reference frame. The agreement between the two sets of corrected velocities is illustrated in Fig.~\ref{fig:velRaw}.

\subsection{Covariance matrix for time-varying CCFs}
\label{sec:covmat}

For calculating instantaneous radial velocities and their precisions from equations~\ref{eq:getvel} and \ref{eq:getvar}, the covariance matrix $\Sigma$ must contain information about the SNR of each observation and the correlations between neighbouring pixels. The covariances between different pixels in the time-series of cross-correlation functions fall into two categories.

The first is systematic uncertainty arising from temporal variability of the profile shape. The {\bf sample} covariance matrix $\Sigma = {\rm Cov}(\mathbf{R}) = \mathbf{R}^T\cdot\mathbf{R}/m$ 
{\bf estimates} the 
systematic covariances between the columns of $\mathbf{R}$, where each column has $m$ elements, one per row of the CCF time series. These arise from correlations between 
different parts of the line profile.

The second is random measurement error arising from the finite SNR of the original spectrum. The CCF sampling interval is matched to the velocity increment per CCD pixel of the instrument. Although neighbouring pixel values in the original spectrum are statistically independent, interpolation during rebinning and calculation of the CCF introduces correlations between neighbouring CCF samples.  We therefore need to account for the correlations between CCF elements at neighbouring values of velocity.

\begin{figure}
    \centering
    \includegraphics[width=\columnwidth]{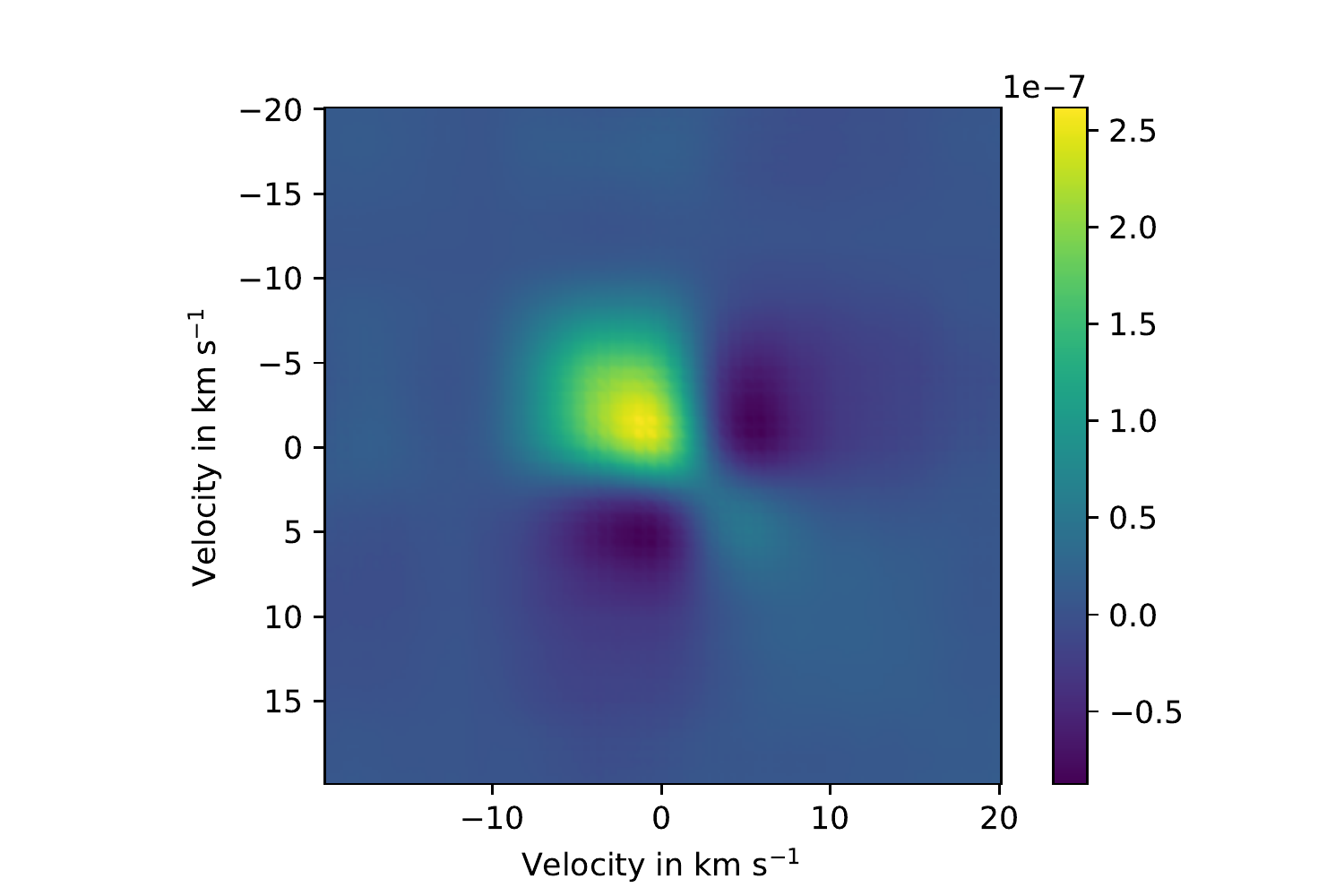}
    \includegraphics[width=\columnwidth]{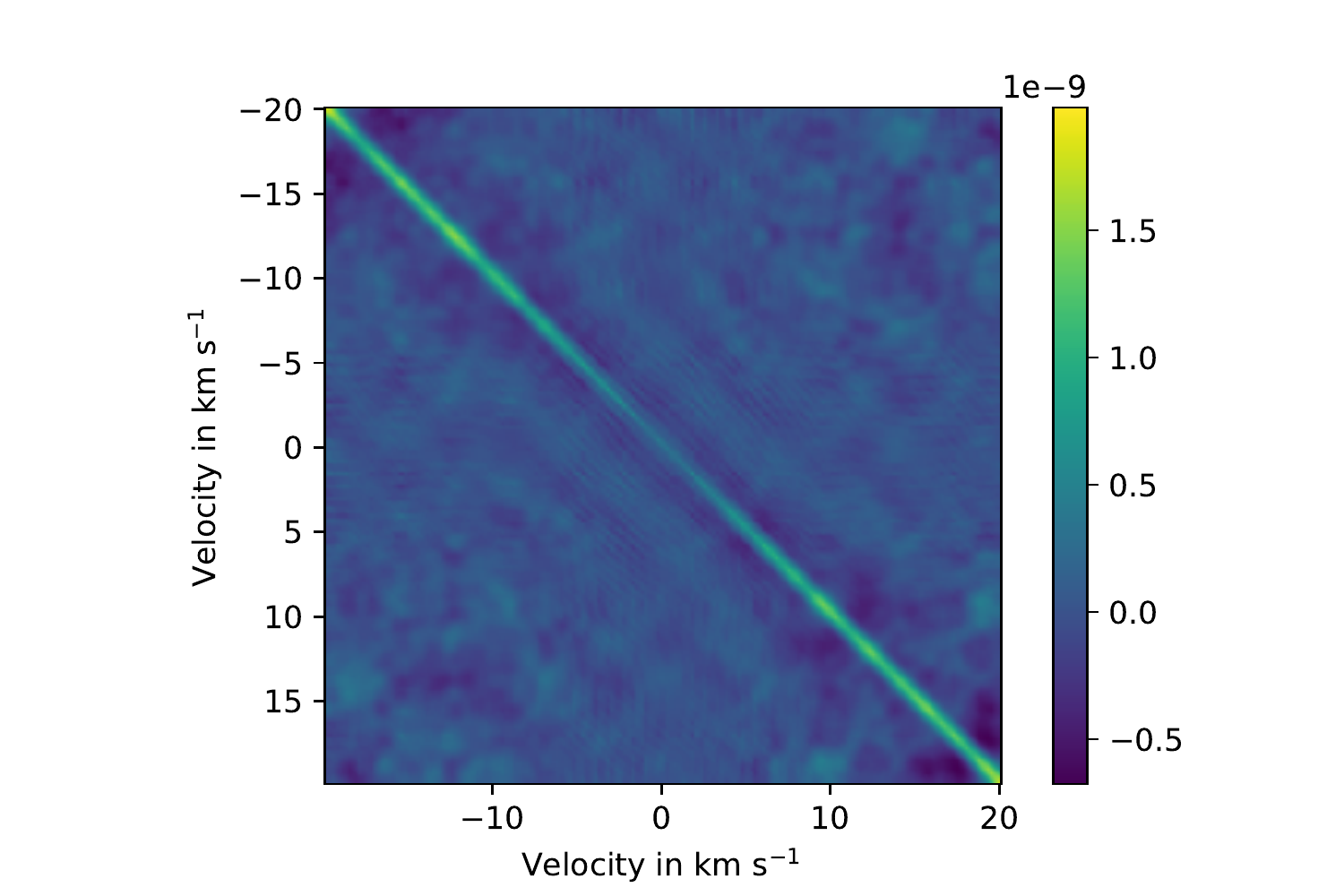}
    \includegraphics[width=\columnwidth]{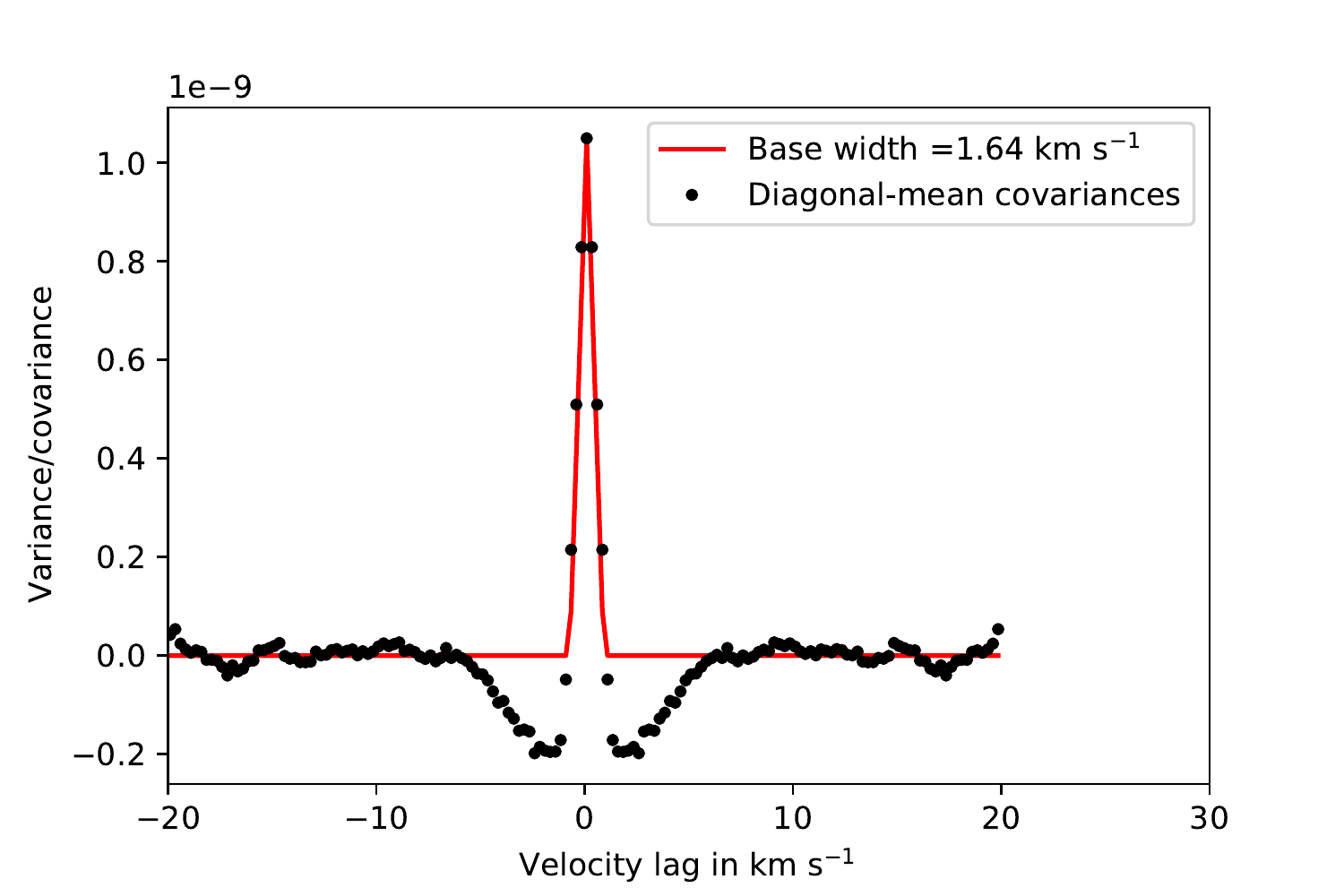}
    \caption{Upper panel: reduced-rank ($k_{\rm max}=10$) covariance matrix of the sequence of CCFs residuals, showing the large-scale covariance due to profile variability. 
    Middle panel: Residual covariance matrix obtained by subtracting the reduced-rank representation from the full covariance matrix. 
    Lower panel: Mean residual covariance along diagonals as a function of horizontal velocity offset from the leading diagonal.
    The diagonal ridge is triangular in cross-section, with base width 1.64 km s$^{-1}$, which is equivalent to two physical pixels on the spectrograph CCD. 
}
    \label{fig:covar}
\end{figure}

The dispersion of HARPS-N gives a near-constant velocity increment per physical CCD pixel in the dispersion direction of 0.82 km s$^{-1}$ \citep{2020arXiv200901945D}. 
The DRS delivers a default CCF that is also sampled in velocity increments of 0.82 km s$^{-1}$, so uncorrelated photon-noise fluctuations in the extracted spectra are spread over about 2 CCF velocity increments after rebinning and interpolation.

The systematic covariances and the instantaneous correlated-noise pattern are unrelated to each other, and enter into the velocity and error calculation in different ways. 
 To estimate the independent systematic uncertainty that affects every observation through profile shape changes, the systematic covariance matrix should be used with equation~\ref{eq:getvar}, separately from the small-scale noise pattern.

The full covariance matrix can be computed via singular-value decomposition of $\mathbf{R}$, as defined in eq.~\ref{eq:ccfSVD}:
\begin{eqnarray}
    \frac{1}{m}\mathbf{R}^T\cdot\mathbf{R}
    &=& \frac{1}{m}(\mathbf{U}_C \cdot {\rm diag}(\mathbf{S}_C)\cdot \mathbf{P}_C^T))^T
    \cdot (\mathbf{U}_C \cdot {\rm diag}(\mathbf{S}_C)\cdot \mathbf{P}_C^T))\nonumber\\
    &=& \frac{1}{m}(\mathbf{P}_C \cdot {\rm diag}(\mathbf{S}_C^2)\cdot \mathbf{P}_C^T).
    \label{eq:covar}
\end{eqnarray}
This approach gives results that are identical to a direct evaluation of the unweighted sample covariance matrix, with a maximum fractional deviation of 1 part in 1000. More importantly, it allows us to calculate a reduced-rank version of the covariance matrix, using only the leading $k_{\rm max}$ principal components. With an appropriate choice of $k_{\rm max}$, a reduced-rank reconstruction of the mean-subtracted CCF residuals
\begin{equation}
    \mathbf{R}_k\equiv\sum_{k=1}^{k_{\rm max}}\mathbf{U}_{C,k} \cdot {\rm diag}(\mathbf{S}_{C,k})\cdot\mathbf{P}_{C,k}^T,
    \label{eq:rk}
\end{equation}
gives a representation of the large-scale temporal covariance pattern that can be written as
\begin{equation}
\frac{1}{m}\mathbf{R}^T_{k_{\rm max}}\cdot \mathbf{R}_{k_{\rm max}} =
\frac{1}{m}(\mathbf{P}_{C,k_{\rm max}} \cdot {\rm diag}(\mathbf{S}_{C,k_{\rm max}}^2)\cdot \mathbf{P}_{C,k_{\rm max}}^T).
\label{eq:covreduced}
\end{equation}
This produces the low-pass filtered covariance matrix shown in the upper panel of Fig.~\ref{fig:covar}.

When this reduced-rank version is subtracted from the full covariance matrix, we obtain the covariance matrix of the remaining high-frequency noise, as shown in the middle panel of Fig.~\ref{fig:covar}. Its strongest feature is a ridge of covariance with an approximately triangular cross-section, whose amplitude depends on the average SNR of the original spectra, and whose width reflects the sampling of the CCF. The peak variance along its diagonal is about 100 times smaller than the peak amplitude of the large-scale covariance pattern. 

The profile of the ridge is seen clearly when the values of the covariance matrix are averaged along diagonals parallel to the leading diagonal, and plotted against velocity lag relative to the leading diagonal, as shown in the bottom panel of Fig.~\ref{fig:covar}. Using a triangular fit
\begin{equation}
    T(v\ | A,v_0,\delta v) = A\  {\rm max}\left(1-\frac{|v-v_0|}{\delta v},0\right)
\end{equation}
to this average profile we obtain an average base half-width
$\delta v = 0.82\, {\rm km s}^{-1}$.
The base width of the ridge perpendicular to the diagonal is therefore about 2 CCF velocity increments, as expected from matching of the CCF sampling interval to the spectrograph resolution element and linear interpolation in the calculation of the CCF. 

We optimise $k_{\rm max}$ to give a clear separation between the large-scale column covariances and the covariances between neighbouring elements in each row arising from oversampling of photon noise. To achieve this we divide the peak value in the bottom panel of Fig.~\ref{fig:covar} by the range of all other diagonal means with lags differing by more than 1 km s$^{-1}$ (slightly more than the CCF sampling interval) in velocity from the leading diagonal. This ratio shows a well-defined peak at $k_{\rm max}=6$, where the separation between the two variance patterns is optimised.

The row variances $\sigma^2_j$ of $\mathbf{R}-\mathbf{R}_{k_{\rm max}}$ reflect the SNR of the individual observations, so we use them as the scale factors $A_j=\sigma^2_j$ of the triangular profile. This leads to the following model for the covariance matrix of the high-frequency noise of an observation made at time $t_j$:
\begin{equation}
    \Sigma_{\xi,\eta}(t_j) \simeq 
 \sigma^2_j\times {\rm max}\left(1-\frac{|v_\xi-v_\eta|}{\delta v},0\right).
    \label{eq:covtrgl}
\end{equation}

This form of the covariance matrix is suitable for calculating the velocities and their precision. The resulting velocities differ from those obtained assuming spatially-uncorrelated noise only at the 15 cm s$^{-1}$ level. Their precision also depends, however, on the scatter introduced by profile-shape changes. When fitting an orbit to the velocities, it is therefore more appropriate to use velocity variances calculated with eq.~\ref{eq:getvar} with a covariance matrix that also includes the reduced-rank model of the covariances arising from the time-varying shape of the CCF:
\begin{equation}
    \Sigma_{\xi,\eta}(t_j) \simeq 
 \sigma^2_j\times {\rm max}\left(1-\frac{|v_\xi-v_\eta|}{\delta v},0\right)
 + \frac{1}{m}\mathbf{R}^T_{k_{\rm max}}\cdot \mathbf{R}_{k_{\rm max}}.
    \label{eq:covmatrix}
\end{equation}

\subsection{Comparison with DRS velocities}

We use this covariance matrix with equations~\ref{eq:getvel} and \ref{eq:getvar} to estimate the radial velocities $\mathbf{\delta v}(t_j)$ and their variances due to photon noise and profile-shape changes.

\begin{figure}
\includegraphics[width=\columnwidth]{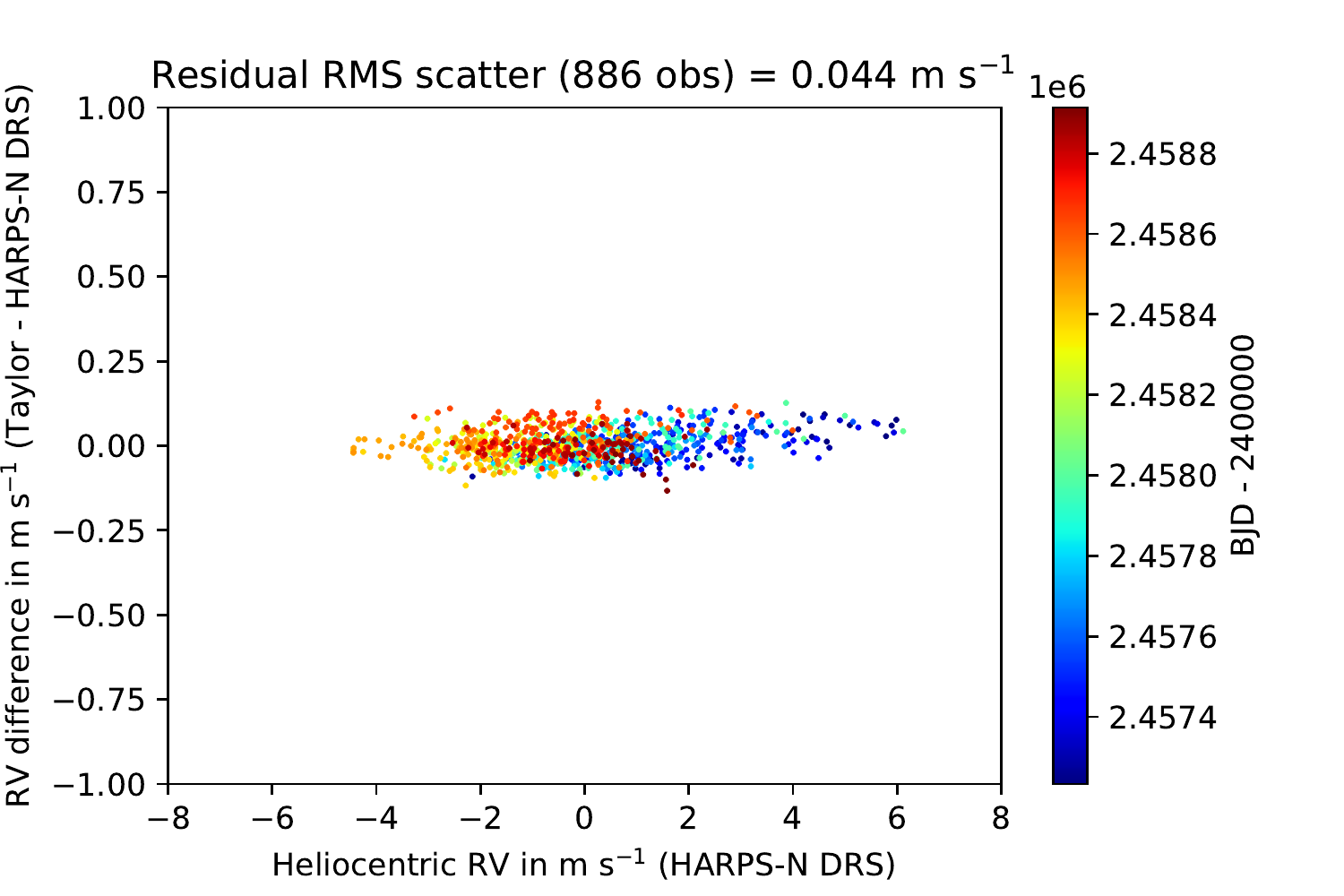}
    \caption{Apparent radial velocities derived from the CCF time series in the heliocentric frame via the Taylor approximation (equation~\ref{eq:getvel}) versus apparent radial velocities from the HARPS-N DRS after applying the barycentric to heliocentric correction. The mean values of both sets of velocities have been subtracted. The standard deviation of the Taylor-series velocities is 2.287 ms $^{-1}$. The two estimators of radial velocity are highly correlated and have slope near unity.  The points are colour coded with BJD-2400000.0. The RMS scatter in the differences between the two sets of measured velocities is 0.044~m~s$^{-1}$.  
}
\label{fig:velRaw}
\end{figure}

In Fig.~\ref{fig:velRaw} we show that radial velocities measured using this method are almost identical to those reported by the HARPS-N DRS. The overall RMS scatter in the differences between the two estimates of the velocity is 0.044~m~s$^{-1}$. 

The formal errors derived from equation~\ref{eq:getvar} are typically 0.11~m~s$^{-1}$, though the dataset includes a small number of days of sparse data yielding uncertainties greater than 0.2~m~s$^{-1}$. Such small error estimates are not surprising. The formal uncertainties computed for individual observations by the DRS indicate an average of 0.24~m~s$^{-1}$ photon noise per exposure. We average many such exposures per day, so the photon noise is insignificant in comparison to the systematic uncertainties arising from profile-shape variations, zero-point errors and calibration drift. 

The independent systematic variances arising from large-scale profile-shape changes are re-computed from equation~\ref{eq:getvar} using the reduced-rank covariance matrix $\Sigma = {\rm Cov}(\mathbf{R}_{k_{\rm max}})$ from equations~\ref{eq:covar} and \ref{eq:rk}. The resulting systematic error per observation is $0.82\pm 0.017$~m~s$^{-1}$, which is very close to the RMS amplitude of the scatter in Fig.~\ref{fig:velHel}.

\section{{\sc scalpels} algorithm}

\subsection{Simple blind search}
\label{sec:appblind}

\textcolor{black}{
{\bf Input:} $\mathbf{C}$ (2D CCF timeseries), $\mathbf{x}$ (CCF array velocities), $\mathbf{v}_{\rm obs}$ (RV measurements), $\sigma(\mathbf{v}_{\rm obs})$ (RV error estimates).
\begin{enumerate}
    \item Compute ACF with eq.~\ref{eq:acfRoll} and normalize.
    \item Compute SV decomposition of ACF with eq.~\ref{eq:acfSVD}.
    \item Perform MAD clip at appropriate threshold on columns of $\mathbf{U}_A$.
    \item Recompute SVD of surviving ACF rows.
    \item Withhold each row of ACF in turn, perform SVD, reconstruct corresponding row of $\mathbf{\hat{U}}_A$ (eqs.~\ref{eq:a_loocv}, \ref{eq:u_loocv}).
    \item Keep columns of $\mathbf{U}_A$ s.t. ${\rm MAD}(\mathbf{U}^T_k-\hat{\mathbf{U}}^T_k)/{\rm MAD}(\hat{\mathbf{U}}^T_k)\ll 1$.
    \item Compute $\hat{\mathbf{\alpha}}$ (eq.~\ref{eq:alpha}).
    \item Rearrange elements of $\hat{\mathbf{\alpha}}$ and columns of $\mathbf{U}_A$ in order of descending $|\delta\chi^2(\mathbf{v}_\perp)|$ (Sect~\ref{sec:rankreduce}).
    \item Select number of basis vectors to keep (Sect~\ref{sec:rankreduce}).
    \item Compute $\mathbf{v}_\parallel$ (eq.\ref{eq:acfRVproj}) in reduced-rank basis.
    \item Compute $\mathbf{v}_\perp=\mathbf{v}_{\rm obs}-\mathbf{v}_\parallel$.
\end{enumerate}
{\bf Return:} $\mathbf{v}_\parallel$ (shape-driven RV), $\mathbf{v}_\perp$ (shift-driven RV).}

\subsection{Simultaneous sinusoidal fit}
\label{sec:appsimul}

\textcolor{black}{
{\bf Input:} $\mathbf{C}$ (2D CCF timeseries), $\mathbf{x}$ (CCF array velocities), $\mathbf{v}_{\rm obs}$ (RV measurements), $\sigma(\mathbf{v}_{\rm obs})$ (RV error estimates), $\{\omega_1,\cdots,\omega_n\}$ (orbital frequencies for $n$ planets).}

\textcolor{black}{Perform steps (i) - (vii) above and see Sect.~\ref{sec:fitsimul}.
\begin{enumerate}
    \item Compute $\mathbf{F} = \{\cos\omega_1 t_j, \sin\omega_1 t_j, \cdots, \cos\omega_n t_j, \sin\omega_n t_j\}$.
    \item Compute reduced-rank covariance matrix (eq.~\ref{eq:covreduced}).
    \item Compute row variances $\sigma^2_j$ of $\mathbf{R}-\mathbf{R}_{k_{\rm max}}$ (Appendix~\ref{sec:CCFexpansion}).
    \item Compute model of full covariance matrix (eq.~\ref{eq:covmatrix}).
    \item Compute $\mathbf{C'}$ and  and ${\rm Var}(\mathbf{v}(t_j))$ (Eq.~\ref{eq:getvar}).
    \item Construct $\Sigma = {\rm Diag}({\rm Var}(\mathbf{v}(t_j)))$.
    \item Construct $\mathbf{P}_\perp=(\mathbf{I}-\mathbf{U}_A\cdot\mathbf{U}_A^T)$ in reduced-rank basis.
    \item Compute $\mathbf{v}_\perp = \mathbf{P}_\perp\cdot\mathbf{v}_{\rm obs}$.
    \item Compute $\mathbf{F}_\perp = \mathbf{P}_\perp\cdot\mathbf{F}$.
    \item Solve eq.~\ref{eq:solvetheta} to obtain $\mathbf{\theta}_{\rm orb}$. 
    \item Compute ${\rm Var}(\mathbf{\theta}_{\rm orb})=1/{\rm Diag}(\mathbf{F}_\perp^T \cdot\mathbf{\Sigma}^{-1}\cdot\mathbf{F}_\perp)$.
    \item Compute $\mathbf{v}_{\rm orb} = \mathbf{F}\cdot\mathbf{\theta}_{\rm orb}$.
    \item Compute $\mathbf{v}_{\rm resid} = \mathbf{v}_\perp-\mathbf{F}_\perp.\theta_{\rm orb})$.
    \item Compute $\mathbf{v}_\parallel=\mathbf{v}_{\rm obs}-\mathbf{v}_\perp$.
\end{enumerate}
{\bf Return:} RV amplitudes and variances, $\mathbf{v}_\parallel$, $\mathbf{v}_\perp$.}

\bigskip
\bigskip
\bigskip
\bigskip
{\it\noindent
$^{1}$SUPA School of Physics and Astronomy, University of St Andrews, North Haugh, St Andrews KY16 9SS, UK\\
$^{2}$Center for Exoplanets and Habitable Worlds,  525 Davey Laboratory, The Pennsylvania State University, University Park, PA 16803, USA\\
$^{3}$Department of Astronomy and Astrophysics,  525 Davey Laboratory,  The Pennsylvania State University, University Park, PA 16803, USA\\
$^{4}$Institute for Computational and Data Sciences, The Pennsylvania State University, University Park, PA 16803, USA\\
$^{5}$School of Physics and Astronomy, Tel Aviv University, Tel Aviv 69978, Israel\\
$^{6}$Department of Physics, University of Oxford, Keble Road, Oxford, OX1 3RH, UK\\
$^{7}$Observatoire Astronomique de l'Universit\'{e} de G\'en\`eve, 51 Chemin des Maillettes, 1290 Sauverny, Suisse\\
$^{8}$Center for Astrophysics | Harvard \& Smithsonian, 60 Garden Street, Cambridge, MA 01238, USA\\
$^{9}$Astrophysics Group, Cavendish Laboratory, Laboratory, University of Cambridge, J.J. Thomson Avenue, Cambridge CB3 0HE, UK\\
$^{10}$DTU Space, National Space Institute, Technical University of Denmark, Elektrovej 328, DK-2800 Kgs. Lyngby, Denmark\\ 
$^{11}$INAF - Fundaci\'on Galileo Galilei, Rambla Jos\'e Ana Fernandez P\'erez 7, E-38712 Bre\~na Baja, Tenerife, Spain\\
$^{12}$INAF - Osservatorio Astronomico di Palermo, Piazza del Parlamento 1, 90134 Palermo, Italy\\
$^{13}$INAF - Osservatorio Astronomico di Cagliari, via della Scienza 5, 09047, Selargius, Italy\\
$^{14}$Dip. di Fisica e Astronomia Galileo Galilei - Universit\`a di Padova, Vicolo dell'Osservatorio 2, 35122, Padova, Italy\\
$^{15}$INAF - Osservatorio Astrofisico di Torino, via Osservatorio 20, 10025 Pino Torinese, Italy\\
$^{16}$Astrophysics Research Centre, School of Mathematics and Physics, Queen's University Belfast, University Road, Belfast, BT7 1NN, UK\\
$^{17}$Kavli Institute for Cosmology, University of Cambridge, Madingley Road, Cambridge CB3 0HA, UK\\
$^{18}$Astrophysics Group, University of Exeter, Exeter EX4 2QL, UK\\
$^{19}$Physics Department, University of Warwick, Coventry CV4 7AL, UK\\
$^{20}$Israel Institute for Advanced Studies, The Hebrew University of Jerusalem Edmond J. Safra Campus, Givat Ram, Jerusalem, Israel\\
$^\dagger$NASA Sagan Fellow 
}

\bsp	
\label{lastpage}
\end{document}